# Role of porosity and diffusion coefficient in porous electrode used in supercapacitors- Correlating theoretical and experimental studies

*Puja De[1], Joyanti Halder[1], Chinmayee Chowde Gowda[2], Sakshi Kansal[3], Surbhi Priya[3], Satvik Anshu[3], Ananya Chowdhury[1], Debabrata Mandal[2], Sudipta Biswas[1], Brajesh Kumar Dubey[4] and Amreesh Chandra[1, 2, 3]\**

[1]Department of Physics, [2]School of Nano Science and Technology, [3]School of Energy Science and Engineering, [4]Department of Civil Engineering, Indian Institute of Technology Kharagpur, Kharagpur, India-7213202.

*E-mail: achandra@phy.iitkgp.ac.in

## Abstract:

Porous electrodes are fast emerging as essential components for next generation supercapacitors. Using porous structures of $Co_3O_4$, $Mn_3O_4$, $\alpha\text{-}Fe_2O_3$, and carbon, their advantages over the solid counterpart is unequivocally established. The improved performance in porous architecture is linked to the enhanced active specific surface and direct channels leading to improved electrolyte interaction with the redox active sites. A theoretical model utilizing the Fick's law is proposed, that can consistently explain the experimental data. The porous structures exhibit ~50-80% increment in specific capacitance, along with high rate capabilities and excellent cycling stability due to the higher diffusion coefficients.

# 1. Introduction

Electrochemical energy storage devices viz., a mobile-ion based secondary batteries and supercapacitors, are amongst the most investigated technologies, which are associated with cleaner energy utilization. The massive progress in these systems is directly linked to the large scale integration of nanomaterial in such devices[1] [2]. Till a few years back, the main focus was to develop nanomaterials with higher specific surface area[3] [4]. It was mostly believed that the higher is the surface area, higher would be the specific capacity of these materials. Since the discovery of graphene and two-dimensional structures like MXenes, it is clear that there are other factors, which drive the enhancement in the storage capacity[5] [6]. For example, layered rGO, $gC_3N_4$, GO, etc., which have lower specific surface area than most of the reported activated carbons, are able to deliver much high specific capacitance[7] [8] [9]. Similar features have been observed in functional metal oxides based 2D- and 3D materials. Here also, the exfoliated 2D structures have been able to deliver much higher specific capacitance than their 2D counter parts, that have higher surface area[10] [11]. Therefore, it is clear that the role of higher specific area of nanomaterials in driving enhanced electrochemical performance is overestimated. This is because, once a binder, conducting component, and active material mixed film is fabricated, the effective surface area is appreciably suppressed. It has been shown that the major additional factors, that lead to enhanced performance are: porosity, ion-channel size, pore volume, interlayer distance, termination elements, etc. [12] [13] [14] [8] [15]. Hence, there has been a paradigm shift in focus, where research is mostly being directed towards the development of porous structures of the various particle morphologies that were being routinely used in electrode materials.

The major advantages of lighter but stable porous electrodes are linked to higher interaction of electrolyte with the active material's surface. This leads to enhanced electrochemical reactions, shorter diffusion lengths and reduced Ohmic, polarization or concentration resistances[16]. This brings appreciable improvement in energy and power density values of the devices. Pseudocapacitive transition metal oxides serve as admirable supercapacitor electrodes. Nonetheless, the poor capacitance and low energy density impose restrictions on their practical applications[17]. According to the working principle of pseudo-capacitors, electrode materials and electrolytes should synergistically combine. The electrolyte ions should transport fast in and out of the bulk of electrode and at the interface between electrode and electrolyte in order to achieve excellent electrochemical properties.

In this paper, using a large number of metal oxides and carbon, the advantages of using porous electrodes in pseudocapacitors and electric double layer (EDL) is unequivocally established. Various morphologies of M-oxides (M= Co, Mn, Fe, Ni, Cu etc.) are generally reported in literature[17] [18] [19] [20] [21] [22] [23]. But a comprehensive report dealing with their porous or hierarchical nanostructures remains elusive. Here, various synthesis protocols, that can be used to easily obtain porous structures, with tunable electrochemical response are discussed. The advantages of these morphologies is proven by comparing the performance with their corresponding solid structures. The electrochemical characterization of the synthesized porous materials show a superior performance, with ~50-80% increment in specific capacitance, in comparison to their solid morphologies. Using a mathematical model, it is also shown that the role of diffusion coefficient becomes critical in enhancing the response characteristics of porous structures. Therefore, the usefulness of porous materials for large scale industrial use is proposed.

## 2. Experimental section

### 2.1 Materials

#### 2.1.1 Synthesis of M-oxides (M= Co, Mn, Fe), and carbon

Cobalt (II) chloride hexahydrate ($CoCl_2.6H_2O$), manganese (II) acetate ($C_4H_6MnO_4$), urea ($CH_4N_2O$), hexadecyltrimethyl-ammonium bromide (CTAB), iron (II) sulfate heptahydrate ($FeSO_4.7H_2O$), oxalic acid ($C_2H_2O_4$), sodium dihydrogen phosphate ($NaH_2PO_4$), and dimethyl formamide (DMF) were purchased from Merck Industries Pvt. Ltd (India). All the precursors were used without further purification.

(i) *Solid and porous nanostructure of $Co_3O_4$*

Porous nanostructures of $Co_3O_4$ were prepared using a low temperature hydrothermal protocol[24]. Initially, $CoCl_2.6H_2O$ and urea were mixed in DI water with a molar ratio of 2:1 and stirred for 30 min. The resultant solution was left undisturbed for ~12 h. Subsequently, the solution was transferred to a sealed borosilicate glass bottle. It was heated in an oven at 120 ºC for 8 h, to obtain a pink precipitate. The precipitate was washed with DI water and ethanol several times to remove the spare urea. The resultant filtrate was vacuum dried at 80 ºC for 12 h. The dried powder was calcined at 500 ºC for 3 h in air, which resulted in the desired final product. The schematic of the synthesis protocol is given in the SI [**Scheme S1(a)**]. The solid nanostructure of $Co_3O_4$ was synthesized by co-precipitaton method[25] as described in the SI.

(ii) *Solid and porous nanostructure of $Mn_3O_4$*

To synthesize porous structures of $Mn_3O_4$ using a hydrothermal method, 0.2 g of hexadecyl trimethyl-ammonium bromide (CTAB) was dissolved in DMF. Subsequently, 0.01 M manganese acetate was added under constant stirring for 1 h. 0.6 g of urea was then added and the solution was poured into a 300 mL Teflon stainless steel autoclave. This was heated at 140 °C for 4 h. The precipitate was washed several times with DI water and ethanol. Finally, the washed precipitate was air dried at 60 °C for 8 h. The schematic of particle growth mechanism is presented in **Scheme S1(b)**. Solid nanostructure of $Mn_3O_4$ was synthesized by following a typical co-precipitation method. The details of the synthesis procedure is provided in the SI.

(iii) *Solid and porous structure of $Fe_2O_3$:*

For obtaining porous $Fe_2O_3$ structures, 0.4 M $FeSO_4.7H_2O$ was mixed with an equal amount of 0.4 M oxalic acid and sodium dihydrogen phosphate solutions. The reaction was kept stirring at 70 °C for 30 min. During the reaction, iron (II) oxalate complex formed, which could be confirmed from the change of colourless solution to yellow. The precipitate was washed with distilled water and ethanol, before being vacuum dried at 45 °C for 12 h. The product was calcined at 400 °C in air for 12 h. This led to the formation of porous microrod of $Fe_2O_3$ [see **Scheme S2** for particle growth mechanism]. The synthesis procedure of solid microrods is also given in the SI.

*(iv)   Food waste derived porous carbons:*

The raw food waste (FW) was used to prepare porous carbon. The collected FW was dried in a hot air oven for 24 h and was grounded using a blender to enhance the consistency and diffusion of sub-critical water into the FW during hydrothermal container. The ground FW was sieved using 475-75μ sieves. The portions containing 300-75μm (to maintain uniform particle size) sieved sample was used for hydrothermal carbonization (HTC) experimentation.

The FW was mixed with distilled water in a 1:10 solid/solvent ratio, and the mixture was eventually subjected to HTC reactor for 6 h at 220 °C, before it naturally cooled to room temperature. The final solid product (hydrochar) was collected through vacuum filtration, washed with distilled water and then oven-dried at 100 °C for 12 h. The prepared hydrochar was further activated by pyrolyzing at 900 °C for 2 h (Ants PROSYS, India), with a slow heating rate of 5 °C/min under the $N_2$ atmosphere[26]. The growth mechanism is depicted in **Scheme S3.** For the solid structure, commercially available activated carbon was used.

## 2.2 Materials characterization

The crystalline phases of all the synthesized materials were analyzed by utilizing the X-ray diffraction (XRD) studies. The diffraction data was collected using a Rigaku Analytical diffractometer with Cu-Kα incident radiation (λ = 0.15406 nm). The morphologies of the synthesized particles were investigated using scanning electron microscopy (SEM CARL ZEISS SUPRA 40) and transmission electron microscopy (TEM FEI-TECHNAI G220S-Twin). The pore size distribution and specific surface area of synthesized materials were tested by the Brunauer-Emmett-Teller (BET) method, using Quantachrome Novatouch surface area and pore size analyser.

## 2.3 Electrochemical characterization

The electrochemical tests, such as cyclic voltammetry (CV), galvanostatic charge-discharge (GCD) of the synthesized materials, were performed using a Metrohm Autolab (PGSTAT302N) potentiostat. The measurements were carried out in 3-electrode cells, using a compatible aqueous electrolyte. Additionally, the three-electrode cell comprised of an Ag/AgCl (in 3 M KCl) as the reference electrode, a platinum rod based counter electrode, and a working electrode. The working electrodes were prepared by mixing 80 wt% of active material, 10 wt% of activated carbon as a conductive agent, and 10 wt % polyvinylidene fluoride as binder, using acetone as a mixing media. The mixture was ultrasonicated and stirred at 60 °C to obtain a homogeneous slurry. The obtained slurry was drop casted uniformly onto a graphite sheet so as to cover an area of 1 $cm^2$. The typical mass loading of the active electrode materials was kept at ~1 mg $cm^{-2}$. Finally, electrode films were vacuum dried at 80 °C for 12 h. Electrochemical impedance spectroscopy (EIS) measurements were also conducted using a Zehner impedance analyser (ZEHNER ZENNIUM pro with THALES XT Software) in the frequency range of 5 mHz to 1 MHz.

## 2.4 Theoretical model

The flux of ionic substance (J) can be expressed by Fick's law. The concentration of electrolyte ions was evaluated from the expression of Fick's law and utilizing the Laplace transformation technique. Using Nernst's and two parameterized equations; one each for charging and discharging, respectively, diffusion coefficient could be estimated. C++ programming was used to write the codes to fit the two parameterized equations.

## 3. Results and discussions

### 3.1 Morphology and structural analysis

Powder X-ray diffraction (XRD) data was analysed to confirm the phase formation of the synthesised materials. **Figure 1(a-d)** depicts the XRD patterns for $Co_3O_4$, $Mn_3O_4$, $Fe_2O_3$, and carbon-based solid and porous powders, respectively. No impurity peaks were observed in any of the profiles, indicating the formation of single-phase materials. The diffraction profiles could be indexed using the relevant JCPDS cards (respective number is mentioned in the figure). The calculated lattice parameters for each of the materials are tabulated in **Table 1.**

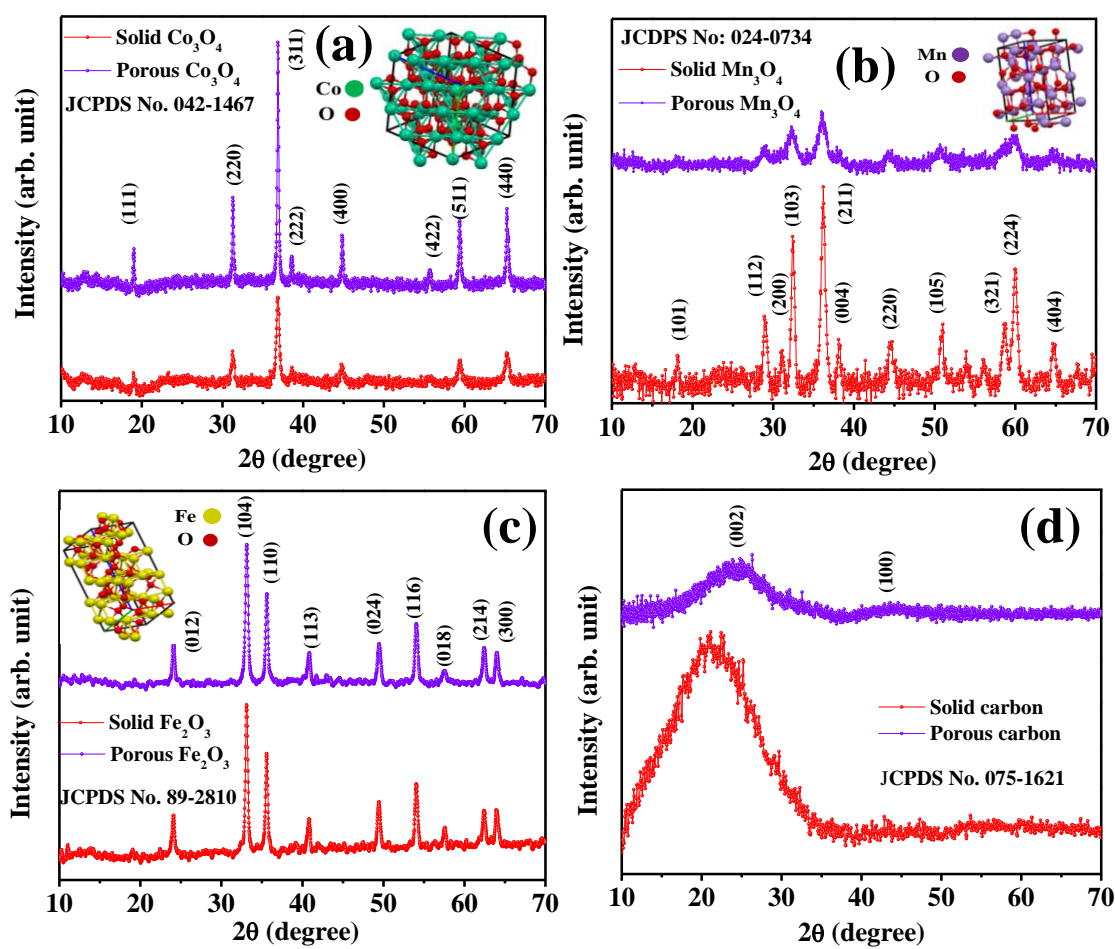

**Figure 1** XRD profiles of solid and porous particles of (a) $Co_3O_4$ (b) $Mn_3O_4$ (c) $Fe_2O_3$ and (d) carbon.

These clearly indicated that the unit cell parameters or the unit cell, for a given metal oxide did not vary significantly with the particle morphology. Hence, if there are changes in electrochemical performance, they can be mostly linked to the variation in the particle shape and sizes.

**Table 1:** Calculated lattice parameters values from the XRD profile

| Material | System name | a | b | c | α | β | γ | Cell volume | Grain size |
|---|---|---|---|---|---|---|---|---|---|
| | | (Å) | | | (degree) | | | ($Å^3$) | (nm) |
| $Co_3O_4$ solid | Monoclinic | 8.09 | 8.08 | 7.96 | 90 | 89.4 | 90 | 520.1 | 18.4 |
| *$Co_3O_4$ porous* | *Monoclinic* | *8.09* | *8.04* | *8.14* | *90* | *90.5* | *90* | *529.4* | *36.4* |
| $Mn_3O_4$ solid | Monoclinic | 6.06 | 5.06 | 9.18 | 90 | 79.4 | 90 | 276.9 | 13.9 |
| *$Mn_3O_4$ porous* | *Monoclinic* | *5.98* | *5.12* | *9.23* | *90* | *79.5* | *90* | *277.8* | *9.5* |
| $Fe_2O_3$ solid | Triclinic | 5.05 | 5.06 | 13.76 | 90.18 | 89.8 | 120.3 | 303.43 | 24.3 |
| *$Fe_2O_3$ porous* | *Triclinic* | *5.05* | *5.06* | *13.77* | *90.31* | *89.76* | *120.3* | *303.76* | *18.5* |

The scanning electron microscopy (SEM) pictures are shown in **Figure 2**. SEM micrographs revealed that the particle size of the porous materials was smaller than their corresponding solid particles. For example, in porous $Co_3O_4$, particles of 50-75 nm were discernible [**Figure 2(b)**], while, in $Co_3O_4$ solid material, particles of > 200 nm were observed [**Figure 2(a)**]. The SEM micrographs for other materials were also investigated and are depicted in **Fig 2(c-h)**. Porous structures of the materials are generally developed by the arrangements of small units. So, porous structures can grow up to a certain range of radius to maintain their stability. In respect to that, solid particles are more stable in nature and can grow bigger in size than porous counterparts. The importance of these smaller and homogeneous particle size distribution in the porous materials will become evident after the electrochemical analysis. In porous structures, particles were loosely bound together, leading to the appearance of directed channels. The TEM micrographs further support this result as shown in **Figure S1**.

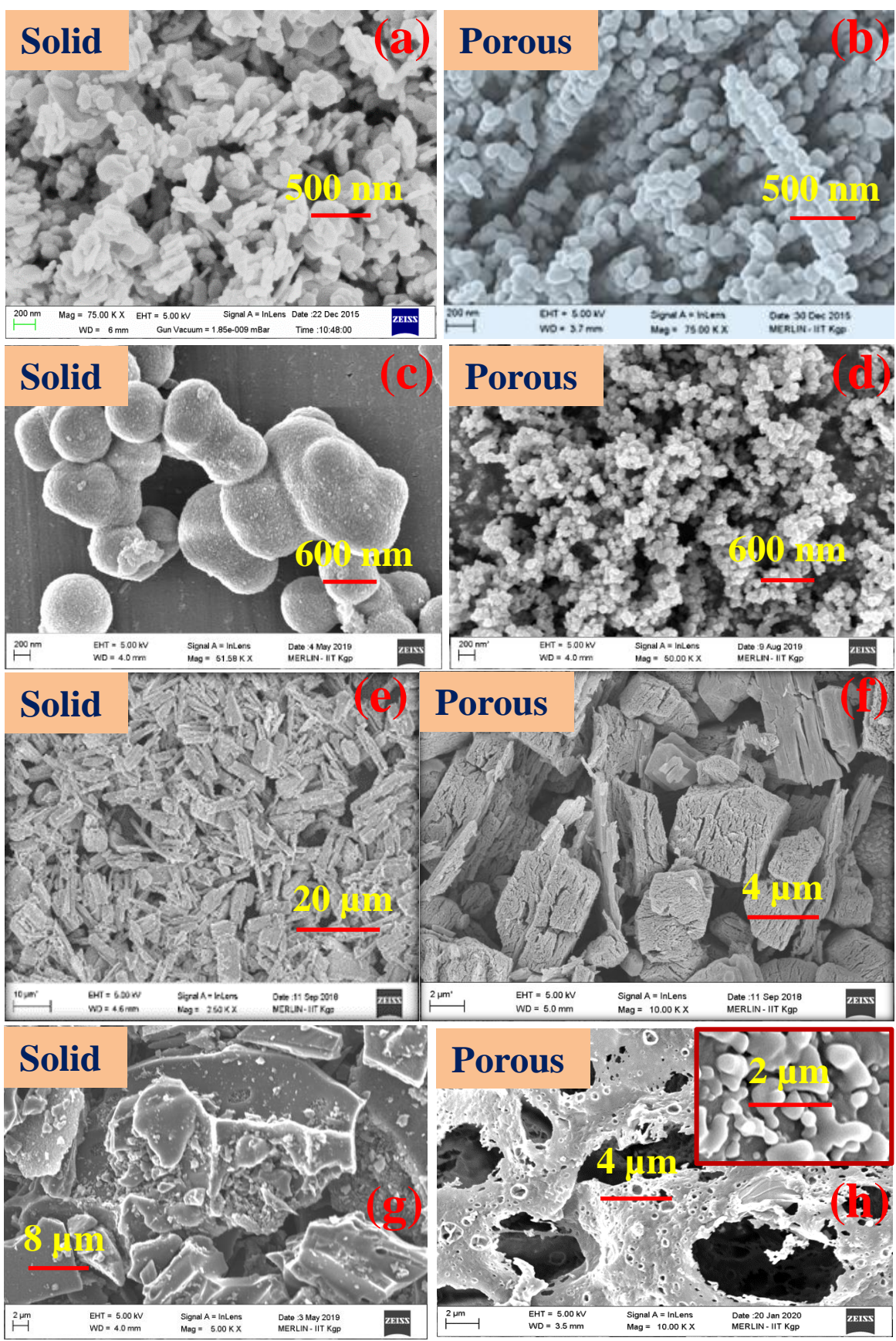

**Figure 2** SEM micrographs of the synthesized solid and porous structure of (a-b) $Co_3O_4$ (c-d) $Mn_3O_4$ (e-f) $Fe_2O_3$ and (g-h) carbon.

In addition to morphological features, the specific surface area plays an important role in determining the capacitive behaviour of these materials. It was determined by collecting and analysing the N$_2$ adsorption-desorption isotherms. **Figure 3** depicts the isotherms for the

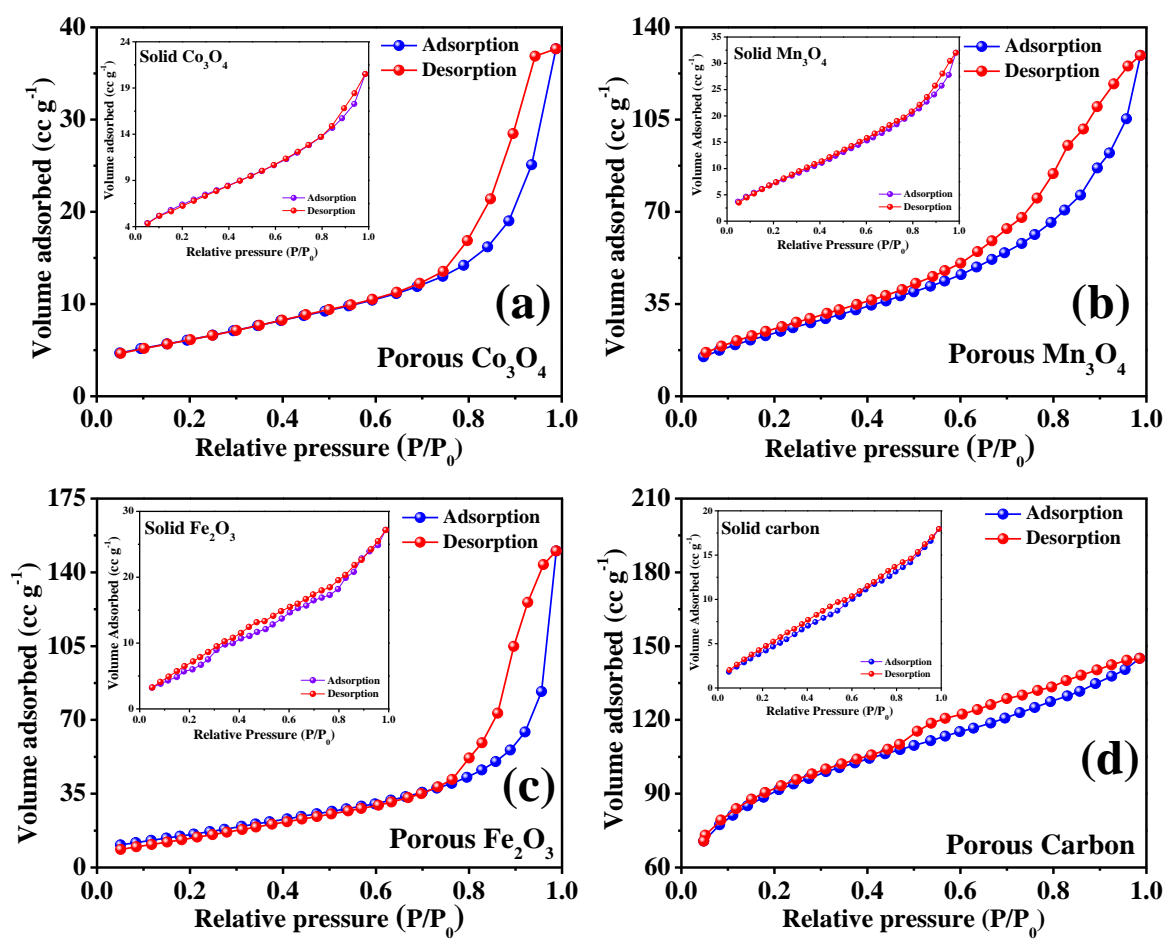

**Figure 3** (a-d) N$_2$ adsorption-desorption isotherms of the synthesized porous structured materials (Inset: corresponding solid structures).

porous and solid structure, respectively, of the investigated materials. The calculated BET specific surface areas for these materials are listed in **Table 2**. Clearly, the specific surface area values of the porous materials were much higher than their solid counterparts.

The pore size and volume were determined by the BJH studies. Mesoporous and microporous structures were observed. The exact values are listed in **Table 2.**

**Table 2:** The calculated surface area and pore radius in solid and porous materials

| Material | Morphology | BET surface area (m$^2$ g$^{-1}$) | Pore radius (nm) |
|---|---|---|---|
| Co$_3$O$_4$ | Solid | 23 | 2.52 |
| | *Porous* | *49* | *7.20* |
| Mn$_3$O$_4$ | Solid | 31 | 1.70 |
| | *Porous* | *91* | *1.82* |
| Fe$_2$O$_3$ | Solid | 30 | 1.58 |
| | *Porous* | *61* | *1.61* |
| Carbon | Solid | 24 | 1.53 |
| | *Porous* | *220* | *1.57* |

**3.2 Electrochemical performances of solid and porous transition metal oxides**

Among the various kinds of aqueous electrolytes (acidic, alkaline and neutral electrolyte), acidic electrolyte (H$_2$SO$_4$) has the highest ionic conductivity. But acidic electrolytes are commonly not suitable for the electrode of metal oxides due to their highest corrosive nature. After the acidic electrolyte, alkaline electrolyte (KOH, NaOH, and LiOH) has the greater conductivity than the neutral electrolyte (KCl, NaCl, K$_2$SO$_4$ and Na$_2$SO$_4$). So, alkaline electrolytes have been utilized extensively in the literature. Mainly, potassium hydroxide (KOH) is most commonly used alkaline aqueous electrolyte, simply because of its higher ionic conductivity (0.54 S cm$^{-1}$), which helps in enhancing the accumulation of charges at the electrode-electrolyte interface. So, here all the measurements were performed in 2 M KOH aqueous alkaline electrolyte. **Figure 4(a-c)** depicts the representative CV curves for solid and porous structures of Co$_3$O$_4$, Fe$_2$O$_3$ and Mn$_3$O$_4$, at a scan rate of 50 mV s$^{-1}$. The CV profiles of porous structure materials enveloped a larger area than their solid counterparts. This indicated towards improved electrochemical properties, as the value of the specific capacitance is directly proportional to the area under the CV curves. The CV profiles of solid and porous structures of TMOs, at various other scan rates, are given in the SI (**Figure S2**). The electrodes showed quasi-rectangular CV curves, with clear reduction-oxidation peaks, indicating the coexistence of pseudocapacitive and as well as the EDL type nature in the materials. The redox peaks could be attributed to the following redox reactions occurring at the respective electrode-electrolyte interfaces:

(i)     $Co_3O_4$:        $Co_3O_4 + H_2O + OH^- \leftrightarrow 3CoOOH + e^-$        (1)

$$CoOOH + OH^- \leftrightarrow CoO_2 + H_2O + e^- \qquad (2)$$

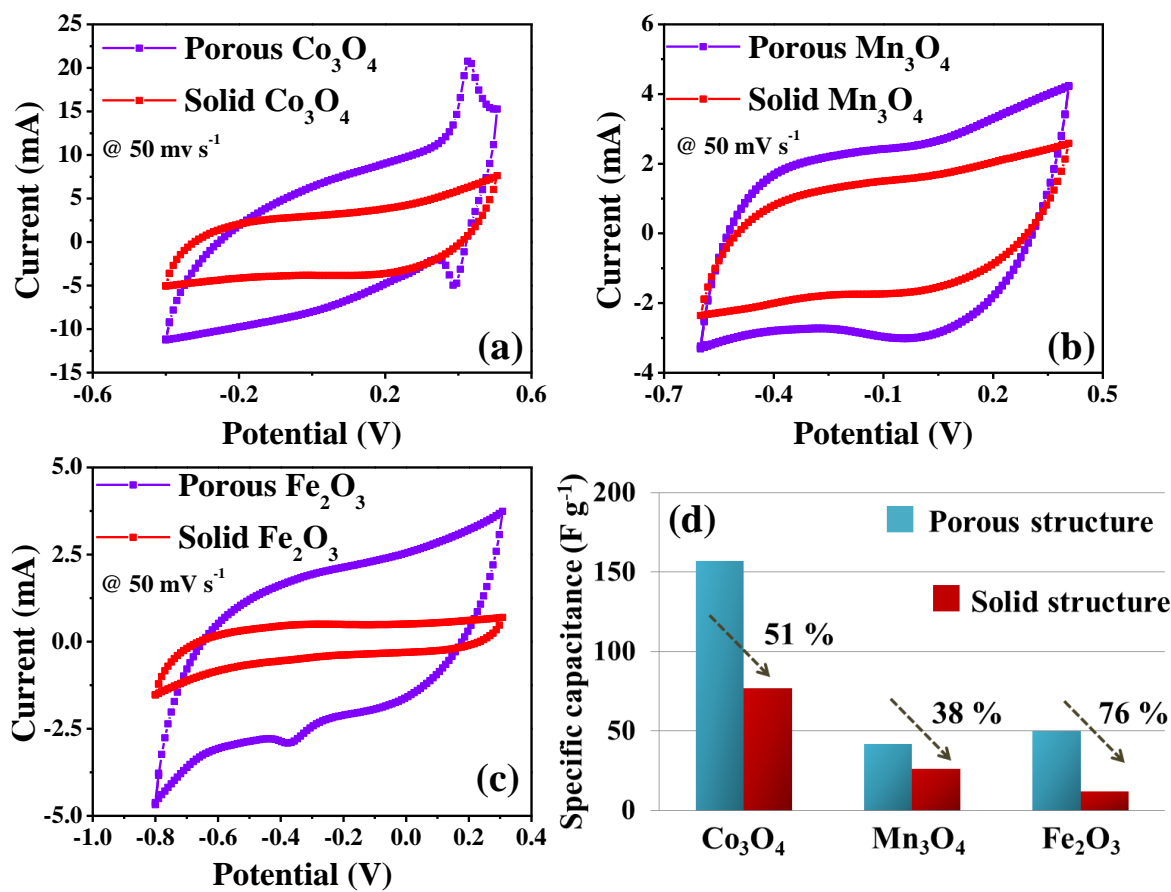

**Figure 4** CV profiles of (a) Co$_3$O$_4$ (b) Mn$_3$O$_4$, and (c) Fe$_2$O$_3$ at 50 mV s$^{-1}$ scan rate. (d) Comparison of specific capacitance between solid and porous morphology of TMOs at 50 mV s$^{-1}$ scan rate.

(ii) **$Mn_3O_4$**: 
$$Mn_3O_4 \rightarrow K_\delta MnO_x \cdot nH_2O \qquad (3)$$

$$K_\delta MnO_x \cdot nH_2O \leftrightarrow MnO_x \cdot nH_2O + \delta K^+ + \delta e^- \qquad (4)$$

(iii) **$Fe_2O_3$**: 
$$Fe_2O_3 + 2e^- + 3H_2O \leftrightarrow 2Fe(OH)_2 + 2OH^- \qquad (5)$$

$$Fe(OH)_2 + 2OH^- \leftrightarrow FeOOH + H_2O + e^- \qquad (6)$$

Interestingly, it was noticed that the redox peaks were more intense in the porous electrodes. As suggested by $N_2$ adsorption-desorption isotherms and microscopic analysis, the porous materials have a higher specific surface area and directed channels. These features could lead to an increased diffusion and transmission of the electrolyte into the porous electrode, resulting in intense redox peaks and higher specific capacitance. Specific capacitance values obtained in these materials are given in **Table 3**. The formula (Equation S1) to calculate specific capacitance from the CV profile is mentioned in SI. At 50 mV s$^{-1}$ scan rate, there was 51, 38, and 76% decrement in the specific capacitance value, on shifting from porous to solid morphologies of $Co_3O_4$, $Mn_3O_4$, and $Fe_2O_3$, respectively. This could be attributed to much reduced interaction between the electrolyte ions and electrode surface. With increasing scan rates, the specific capacitance value also gradually decreased, as shown in **Figure S3.** This happens owing to the fact that the electrolyte ions, at faster scan rates, do not have sufficient time for diffusion into the electrode.

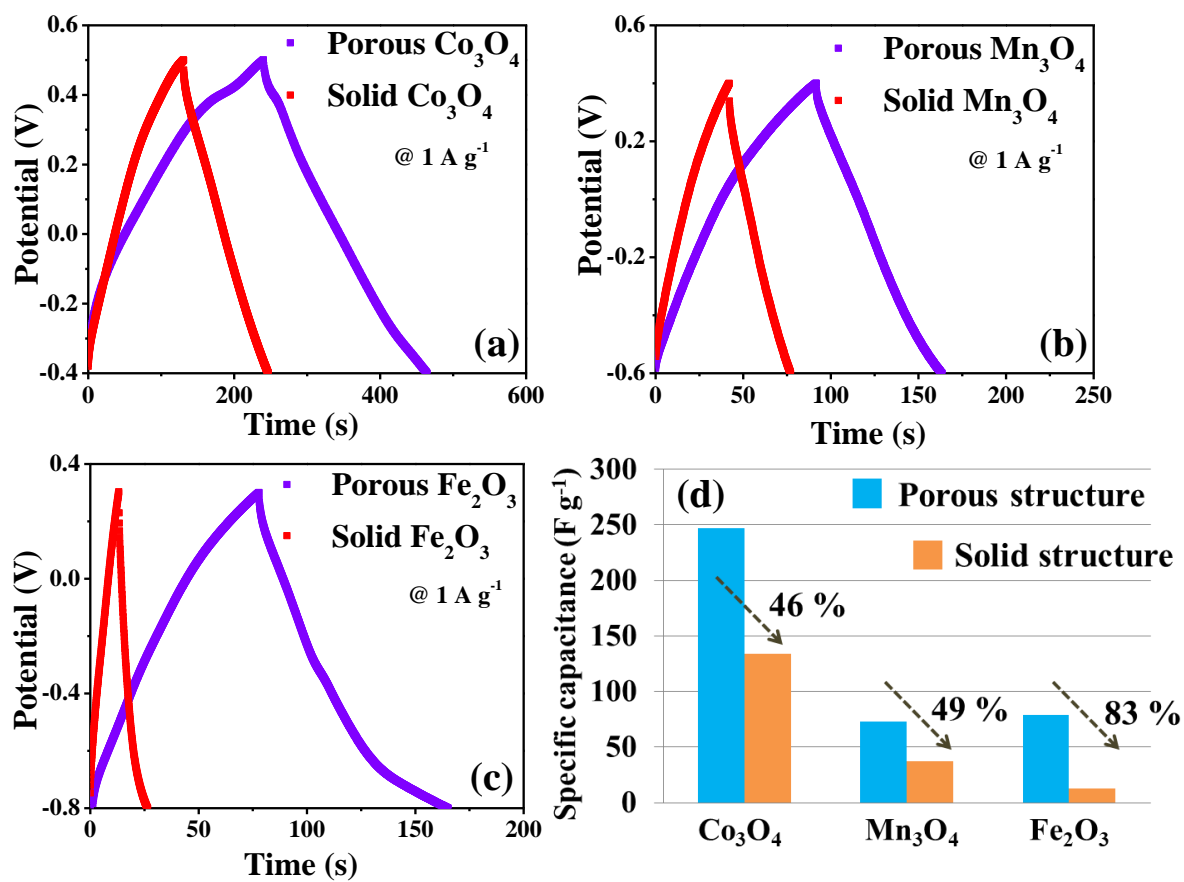

**Figure 5** CD profile of (a) $Co_3O_4$ (b) $Mn_3O_4$, and (c) $Fe_2O_3$ at 1 A g$^{-1}$ current density. (d) Comparison of specific capacitance between solid and porous morphologies at 1 A g$^{-1}$ current density.

To further reaffirm the higher electrochemical behaviour of porous electrodes, the materials were further characterized using galvanostatic charge-discharge studies. **Figure 5(a-c)** presents the observed GCD profiles at 1 A g$^{-1}$ current density, in both solid and porous particles. The results further proved the superiority of porous structures. It was observed that, for porous electrodes, the discharge curves had two distinct regions. A linear region associated with the EDLC nature and a non-linear region, which is normally linked with pseudocapacitance. The specific capacitances were estimated from the GCD profiles at 1 A g$^{-1}$ current density using the formula given in the SI (Equation S2). The specific capacitance values are given in **Table 3**. The GCD profiles, at current densities ranging from 1 to 5 A g$^{-1}$, were further performed (**Figure S4**) to investigate the electrochemical rate capabilities at higher scan rates. **Figure S5** gives the specific capacitance of these samples at various current densities. Clearly, the capacitance values decreased with increasing current densities. At higher specific currents, the electron transfer towards the electrode is faster, and hence the increment in potential is higher. Consequently, the electrode has reduced time to stay at a specified voltage and a lower specific capacitance value is observed. The electrochemical results at a higher current density of 5 A g$^{-1}$ revealed that porous architectures of $Co_3O_4$, $Mn_3O_4$, and $Fe_2O_3$ showed excellent specific capacitance retention of 52, 67, and 58% of their maximum value at 1 A g$^{-1}$, whereas the solid architectures retained only 41, 37, and 50%, respectively.

The electrochemical properties of solid and porous structured carbon were also investigated and compared. The detailed CV and charge-discharge profiles are given in SI [see **Figure S6**]. Carbon-based electrode showed ~52% enhancement in the specific capacitance value on moving from solid to porous structure, as given in **Table 3**.

**Table 3:** Calculated specific capacitance in the solid and porous materials using the CV and GCD profiles.

| Material | Morphology | Specific capacitance from CV profile (F g$^{-1}$) at a fixed scan rate | Specific capacitance from GCD profile (F g$^{-1}$) at 1 A g$^{-1}$ current density |
|---|---|---|---|
| $Co_3O_4$ | Solid | 121 F g$^{-1}$ at 5 mV s$^{-1}$ | 134 |
| | *Porous* | *241 F g$^{-1}$ at 5 mV s$^{-1}$* | *247* |
| $Mn_3O_4$ | Solid | 32 F g$^{-1}$ at 5 mV s$^{-1}$ | 37 |
| | *Porous* | *55 F g$^{-1}$ at 5 mV s$^{-1}$* | *73* |
| $Fe_2O_3$ | Solid | 12 F g$^{-1}$ at 50 mV s$^{-1}$ | 13 |
| | *Porous* | *50 F g$^{-1}$ at 50 mV s$^{-1}$* | *79* |
| Carbon | Solid | 84 F g$^{-1}$ at 20 mV s$^{-1}$ | 78 |
| | *Porous* | *128 F g$^{-1}$ at 20 mV s$^{-1}$* | *114* |

To further understand the charge transport kinetics of the materials during the electrochemical response, EIS measurements were performed and the corresponding Nyquist plots are shown in **Figure 6 (a-c)**. The Nyquist plot consisted of two portions i.e. a semi-circular region in the high frequency range and a straight line at the low frequency range. The semicircle in the high-frequency region is attributed to the interfacial charge transfer resistance ($R_{ct}$) formed between the electrolyte and the surface of the electrode, which can be estimated from the diameter of the semicircle [(inset of **Figure 6(a-c)**]. The magnitude of diameter of the semicircles suggested that the porous electrodes had a lower value of $R_{ct}$, compared to their solid counterparts. The straight line in the low-frequency region is attributed to the Warburg diffusion process. The linear curve at ~45° confirmed their capacitive nature and it is noteworthy that porous structured electrodes displayed more vertical EIS curves. This indicated towards a reduced diffusion resistance. The obtained $R_{ct}$ and ESR values are tabulated in **Table 4**.

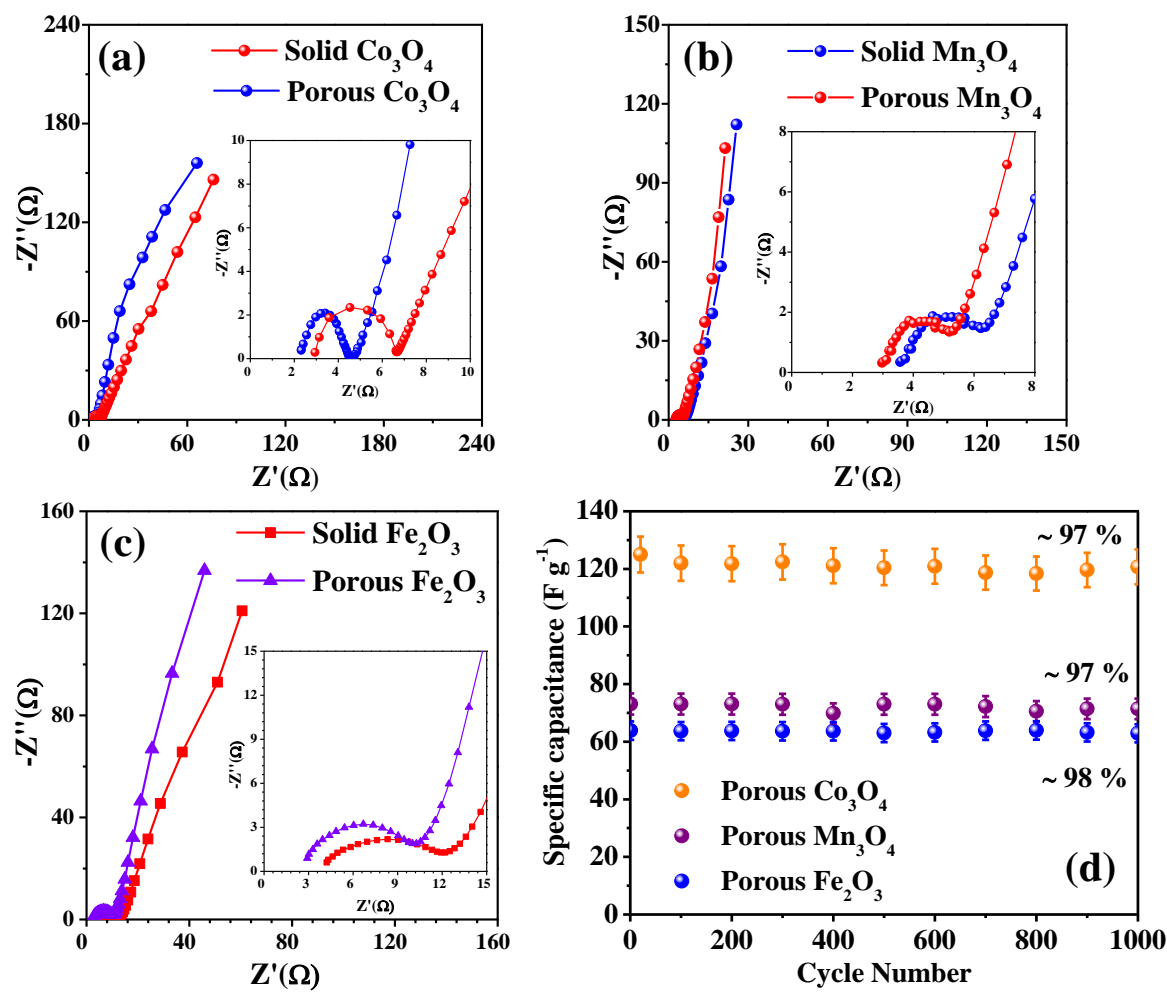

**Figure 6** Nyquist plot of porous and solid morphology of (a) $Co_3O_4$ (b) $Mn_3O_4$ and (c) $Fe_2O_3$. (d) Cycling stability test of porous $Co_3O_4$, $Mn_3O_4$ and $Fe_2O_3$.

Finally, the electrochemical stability of the porous materials was evaluated by repetitive charging and discharging, at a higher current density, for 1000 cycles. The porous materials showed excellent cycling stability, with a capacity retention of ~97-98 %, as shown in **Figure 6(d)**. This stability could be attributed to a limited volume change in porous structures, owing to the presence of the voids within the material.

**Table 4:** Charge transfer ($R_{ct}$) and ESR values of solid and porous structure of $Co_3O_4$, $Mn_3O_4$ and $Fe_2O_3$.

| Material | Solid structure | | Porous structure | |
|---|---|---|---|---|
| | ESR (Ω) | $R_{ct}$ (Ω) | *ESR (Ω)* | *$R_{ct}$ (Ω)* |
| $Co_3O_4$ | 2.92 | 4.52 | *2.03* | *2.22* |
| $Mn_3O_4$ | 3.51 | 3.42 | *2.95* | *3.15* |
| $Fe_2O_3$ | 4.25 | 7.72 | *2.91* | *7.06* |
| Carbon | 1.85 | 3.82 | *1.77* | *3.45* |

The EIS characterization and cyclic stability of food waste derived porous carbon and the solid carbon are presented in SI (**Figure S7**).

The results obtained in the presented porous materials was compared with the published literature. This is elaborated in **Table 5**.

**Table 5:** Comparison of electrochemical results of the porous $Co_3O_4$, $Mn_3O_4$, $Fe_2O_3$ and carbon with some published results on the various morphologies and composites of similar materials.

| Material | Structure | Current density/scan rate | Specific capacitance (F $g^{-1}$) | Cycling stability | Ref. |
|---|---|---|---|---|---|
| $Co_3O_4$ | | | | | |
| $Co_3O_4$ | Porous nanocubes | 0.2 A $g^{-1}$ | 350 | ---- | [27] |
| $Co_3O_4$ | Hollow boxes | 0.5 A $g^{-1}$ | 278 | ---- | [28] |
| $Co_3O_4$/Ni-Co layered double hydroxide | Core-shell | 0.5 A $g^{-1}$ | 318 | 92% after 5000 cycles | [29] |
| $Co_3O_4$ | nanoparticles | 0.2 A $g^{-1}$ | 179.7 | 73.5% after 1000 cycles | [30] |
| Needle-like $Co_3O_4$/graphene | Composite | 0.1 A $g^{-1}$ | 157.7 | 70 % after 4,000 cycles | [31] |
| $Co_3O_4$ | Layered parallel folding | 1 A $g^{-1}$ | 202.5 | ~99% after 1,000 cycles | [32] |
| $Co_3O_4$ | Porous nanorod | 1 A g$^{-1}$ | 247 | ~97 % after 1,000 cycles | **Present work** |
| **$Mn_3O_4$** | | | | | |
| $Mn_3O_4$ | Porous hollow microtube | 1 mA cm$^{-2}$ | 51.7 | 57 % after 2000 cycles | [33] |
| $Mn_3O_4$ | Thin film | 50 mV s$^{-1}$ | 12.3 | 91 % after 2000 cycles | [34] |
| $Mn_3O_4$ | Nanoparticle | 1 A g$^{-1}$ | 55.7 | 81 % after 1200 cycles | [35] |
| $MnOOH@Mn_3O_4$ | Composite | 1 A g$^{-1}$ | 71 | ------ | [36] |
| $Mn_3O_4$ | Hollow nanofiber | 0.3 A g$^{-1}$ | 155 | 99 % after 500 cycles | [37] |
| $Mn_3O_4$ | Porous | 1 A g$^{-1}$ | 73 | 97 % after 1000 cycles | **Present work** |
| **$Fe_2O_3$** | | | | | |
| $Fe_2O_3$ | Nanotube | 2.5 mV s$^{-1}$ | 30 | 92 % after 2,000 cycles | [38] |
| $Fe_2O_3/MnO_2$ | Core-shell | 0.1 A g$^{-1}$ | 159 | 97.4 % after 5000 cycles | [39] |
| $Fe_2O_3$/Ni-foam | Sheets | 0.36 A g$^{-1}$ | 147 | 86 % after 1,000 cycles | [40] |
| $Fe_2O_3$ | Bulk | 1 A g$^{-1}$ | 40 | ---- | [41] |
| $Fe_2O_3$ | Rod | 1 A g$^{-1}$ | 80 | 98% after 1,000 cycles | **Present work** |
| **Carbon** | | | | | |
| Bare carbon | Porous | 1 mA cm$^{-2}$ | 92 F g$^{-1}$ | 97.8 % after 1000 cycles | [42] |
| Porous carbon | nanofibers | 25 mV s$^{-1}$ | 98.4 F g$^{-1}$ | ------- | [43] |
| CNF | nanofiber | 1 mA cm$^{-2}$ | 60 F g$^{-1}$ | ------- | [44] |
| Rice husk derived carbon | Porous (RHPC) | 1 mA cm$^{-2}$ | 60 F g$^{-1}$ | 78 % after 5000 cycles | [45] |
| Porous carbon aerogel | Hierarchical hole-like | 1 A g$^{-1}$ | 132 F g$^{-1}$ | 93.9% after 5000 cycles | [46] |
| Food waste derived-carbon | Porous | 1 A g$^{-1}$ | 114 F g$^{-1}$ | 96% after 2000 cycles | **Present work** |

**3.3 Theoretical interpretation of the performance enhancement in porous electrodes**

Several factors affect the electrochemical performance of an electrode material. Amongst them, electrolyte-electrode interaction is important in both EDLCs and pseudocapacitors. Porous structures significantly enhance the contact area between the electrode and electrolyte ions. Hence, they are expected to have higher electrochemical performances. The channels facilitate higher ion intercalations and deintercalation. **Scheme 1** highlights the fact of the presence of higher electrochemically active sites in porous structures as compared to their solid counterparts.

In solid structure, the electrolyte ions are confined only at the surface of the electrode, whereas the presence of the larger pores, in porous electrode, provides much higher access for the transmission of electrolyte ions through the material. Hence, the role of the diffusion process becomes important. This can be modelled using the Fick's law[47] of ionic diffusion. According to this law, flux J of an ionic substance, at a position x and time t, is given by:

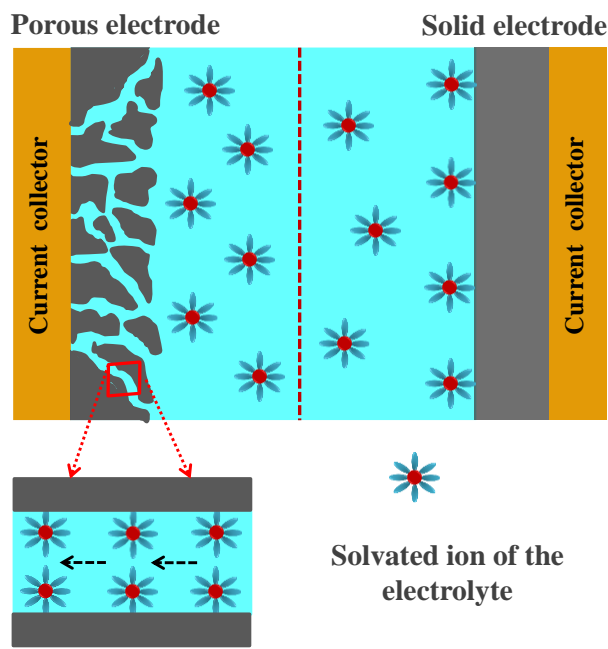

**Scheme 1** Schematic representation of the transportation of electrolytes ions through the solid and porous structured electrode material.

$$J = -D\frac{\partial C}{\partial x} \qquad (7)$$

where, C represents the concentration of the electrolyte ions near the electrode surface and D is the diffusion coefficient of the electrolyte ions. By calculating the concentration of electrolyte ions C, during charging and discharging and substituting it in to the Nernst's equation[48], the potential of the electrode surface can be obtained as:

$$E = E_0 + \frac{RT}{nF}\ln\left(\frac{C_0}{C_R}\right) \qquad (8)$$

where, $E_0$ denotes the reference potential of the electrochemical system. Hence, the substitution of analytical expression of $C_0(0,t), C_R(0,t)$, ($C_0$: concentration of electrolyte ions during oxidation, $C_R$ is that for reduction) gives an expression for the time-varying electrode potential across the solid electrolyte interface (SEI). This leads to two time-dependent equations for the electrode surface (owing to charging and discharging). After parameterization, the two equations can be written:

$$E = a_1^c + b_1^c t^{\frac{1}{2}} + c_1^c t \qquad \text{for charging} \qquad (9)$$

$$E = a_2^d + b_2^d (t^{\frac{1}{2}} - e_2^d (t - t_c)^{\frac{1}{2}}) + c_2^d \left(t^{\frac{1}{2}} - e_2^d (t - t_c)^{\frac{1}{2}}\right)^2 \qquad \text{for discharging} \qquad (10)$$

where, $a_1^c$, $b_1^c$, $c_1^c$, and $a_2^d$, $b_2^d$, $c_2^d$ are the coefficients of charging and discharging, respectively. The system achieves the highest potential ($E_{max}$) during charging, corresponding to the time taken to fully charge ($t_c$). Hence, the charging time can be obtained from the following equation[49]:

$$E_{max} = a_1^c + b_1^c t_c^{\frac{1}{2}} + c_1^c t_c \qquad (11)$$

where $a_1^c$, $b_1^c$ and $c_1^c$ are the values obtained from the fitted data. Similarly, discharge time $t_d$ was evaluated from the equation:

$$E_{min} = a_2^d + b_2^d (t_d^{\frac{1}{2}} - e_2^d (t_d - t_c)^{\frac{1}{2}}) + c_2^d \left(t_d^{\frac{1}{2}} - e_2^d (t_d - t_c)^{\frac{1}{2}}\right)^2 \qquad (12)$$

By subtracting $t_c$ from the value of $t_d$, one can estimate the time taken to discharge a supercapacitor. From the value of charging and discharging coefficients, the diffusion coefficient of electrolyte ions can be easily obtained. For current varying electrochemical cells, the potential across the electrode advances as a function of time. On accumulation of charges, the potential differences at the interface can change with time but this effect is neglected in the calculation, to maintain simplicity[49]. This leads to:

$$\frac{\partial C}{\partial t} = D \frac{\partial^2 C}{\partial x^2} \qquad (13)$$

As one can see, Equation 13 is a 2nd order differential equation, involving time and position. The details of solving the equation has been already reported by our group[49]. Here also, a similar protocol was followed. Solving equation (13) and substituting it in Nernst's equation, electrode potential during charging and discharging can be written as:

$$E \approx E_0 - \frac{RT}{2nF}\ln\left(\frac{D_O}{D_R}\right) - \frac{RT}{nF}\left((i\alpha)t^{\frac{1}{2}} + (i\alpha)^2 \frac{t}{2}\right) \quad \text{for charging} \tag{14}$$

and

$$E \approx E_0 - \frac{RT}{2nF}\ln\left(\frac{D_O}{D_R}\right) + \frac{RT}{nF}\left(-(i\alpha)\left(t^{\frac{1}{2}} - 2(t-t_c)^{\frac{1}{2}}\right) - \frac{(i\alpha)^2}{2}\left(t^{\frac{1}{2}} - 2(t-t_c)^{\frac{1}{2}}\right)^2\right) \tag{15}$$

for discharging

where, E represents the potential of the electrode surface both for charging and discharging. The terms $E_0$, $i$, $D_O$ and $D_R$ stand for the reference potential of the electrochemical system, current density, and the initial and final diffusion coefficient, respectively. $F, R, T$ represent the Faraday number, the universal gas constant, and the temperature of the system, respectively, whereas, $\alpha$ is a constant parameter of the system which take the following value: $\alpha = \left(\frac{2}{nFAC_0^* D_0^{\frac{1}{2}} \pi^{\frac{1}{2}}}\right)$, where A, D$_0$, and $C_0^*$ represent the area of the electrode, the diffusion coefficient, and the concentration of ion at time t = 0. Further, a comparison of Equations 9, 10, 14 and 15 give the values[25]:

$$b_1^c = -\frac{RT}{nF}(i\alpha), \quad c_1^c = -\frac{RT}{2nF}(i\alpha)^2 \quad \text{for charging}$$

$$b_1^d = -\frac{RT}{nF}(i\alpha), \quad c_1^d = -\frac{RT}{2nF}(i\alpha)^2 \quad \text{for discharging}$$

Therefore, $2\frac{c_1^c}{b_1^c} = (i\alpha)$ \hfill (16)

On solving Equation (16), one can obtain the value of $i\alpha$, which is required during the calculation of diffusion coefficients. This was obtained as:

$$i\alpha = \left(\frac{2i}{nFAC_0^* D_0^{\frac{1}{2}} \pi^{\frac{1}{2}}}\right) \tag{17}$$

where i is the applied current density and n is the number of electrons passing the electrode during the reaction.

So, to establish the statement of *higher electrolyte ions diffusion occurs through the porous electrodes*, the values of the diffusion coefficients were calculated from the charging and discharging coefficients; obtained by fitting the CD profiles of the porous and solid particles. **Figure 7(a, c, e)** shows the comparison curve of theoretical and experimental data for porous $Co_3O_4$, $Mn_3O_4$, and $Fe_2O_3$, respectively. All of the GCD curves obtained at 1 A g$^{-1}$ current density were used. Comparison curves for solid structures of metal oxides and solid and porous carbon are given in SI [**Figure S8(a-e)**]. The theoretical curves were in good agreement with the experimental data, with low relative error (<5%) [see **Table 6**]. Like the experimental results, the specific capacitance for each porous material achieved a greater value than their solid structures; the values are tabulated in **Table 6**. This can be explained by the diffusion behaviour of the electrolyte ions. **Figure 7(b, d, f)** illustrates the comparison of diffusion coefficients of electrolyte ions in solid and porous structured $Co_3O_4$, $Mn_3O_4$, and $Fe_2O_3$, respectively. The variation of diffusion coefficient for solid and porous structured carbon is shown in **Figure S8(f)**. The diffusion coefficients of the electrolyte ions were calculated for solid and porous structures at 1 A g$^{-1}$ current density. In each of these three metal oxides, and carbon, the diffusion coefficient increased when we moved from solid to porous structures. The diffusion coefficients in porous and solid structures are given in **Table 6**. The higher D values in porous nanostructures further proved their importance. It could be inferred that there exists a correlation between diffusion coefficient and specific capacitance. Diffusion coefficients depend upon different factors. Amongst them, the morphology of electrode material is critical. Usually, the electrochemical performance increases due to the increase in mobility of the electrolyte ions in to porous structures.

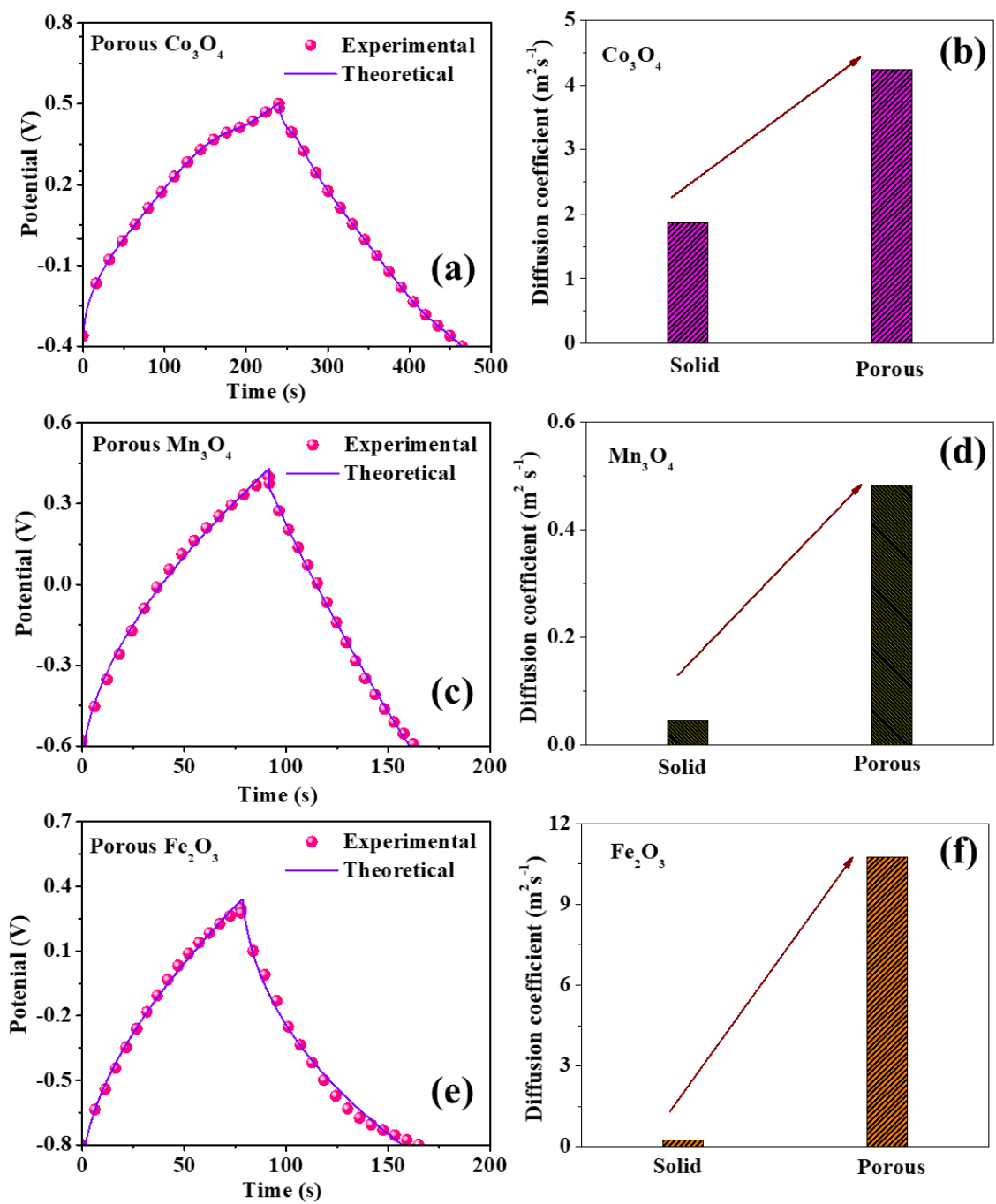

**Figure 7** Comparison between theoretical and experimental data at 1 A g$^{-1}$ current density and diffusion coefficient of (a, b) Co$_3$O$_4$ (c, d) Mn$_3$O$_4$ (e, f) Fe$_2$O$_3$, respectively.

**Table 6:** Theoretical specific capacitance with relative errors and diffusion coefficients values of solid and porous structure of the materials.

| Material | Morphology | Theoretical specific capacitance (F g$^{-1}$) | Relative error (%) | Diffusion coefficients (m$^2$ s$^{-1}$) |
|---|---|---|---|---|
| $Co_3O_4$ | Solid | 135.9 | 1.42 | 1.8744 |
| | *Porous* | *243.0* | *1.62* | *4.244* |
| $Mn_3O_4$ | Solid | 35.2 | 4.81 | 0.0453 |
| | *Porous* | *72.4* | *0.82* | *0.4824* |
| Fe$_2$O$_3$ | Solid | 12.0 | 7.69 | 0.0535 |
| | *Porous* | *78.0* | *1.26* | *10.755* |
| Carbon | Solid | 75.4 | 3.33 | 0.01904 |
| | *Porous* | *104.1* | *3.42* | *0.1153* |

The kinetics of the electrolyte ion changes with the change in porosity of the materials. Each of the porous materials discussed here had mesoporous or microporous nature, with higher pore radius than their solid counterparts. Due to the increase in porosity, diffusion (of electrolyte ions) can occur in several directions. Also, effective diffusion length becomes larger and lead to increase in the diffusion coefficient[50].

So, both the theoretical modelling and experimental analysis demonstrated the improved electrochemical performance of porous structure electrode at a given rate compared to their solid bulk material. There was ~ 46, 49, 83, and 32% decrement in specific capacitance value when the porous structures of the Co$_3$O$_4$, Mn$_3$O$_4$, Fe$_2$O$_3$ and carbon were replaced by their corresponding solid counterpart. The BET results also suggested that the porous materials have greater specific surface area i.e. large area per unit mass, which inturn facilitates enhanced contact area of electrode-electrolyte interface. The larger contact area mean that the current density per unit surface area decreases, which reduces the electrode polarization and drives higher charge transfer at the interface.

The pore in the porous Co$_3$O$_4$, Mn$_3$O$_4$, Fe$_2$O$_3$ and carbon was also large. These would act as an ''ion reservoir" and have ample room, to absorb lattice expansion during cycling. Therefore, the porous materials are able to deliver excellent cycling stability of > 97% after 1,000 cycles. The porous material also showed lower values of ESR and R$_{ct}$. This confirmed the facile charge transfer reactions and much lower diffusion resistance of the electrolyte ions at the interface.

## 4. Conclusions

It is clearly shown that the porous nanostructures of transition metal oxides and carbon have higher specific capacitance values than their solid counterparts. Porous structures of $Co_3O_4$ was able to deliver ~84% increment in specific capacitance value, at 1 A g$^{-1}$ current density, in comparison to its solid counter parts. Other investigated materials also showed similar results. The enhanced performance could be attributed to the higher surface adsorption sites and specific surface area in the porous structures. The porous network facilitates higher ion diffusion and improved electrochemical interactions. The porous materials also showed higher cyclic retention, with lower ESR values. All the experimentally observed data are consistently explained by a theoretical model, which uses Ficks law as the starting point to calculate the extent of ion-diffusion within the material.

## Conflicts of interest

The authors declare no conflicts of interest.

## Acknowledgements

The authors acknowledge the funding received from DST (India) under its MES scheme for the project entitled, "Hierarchically nanostructured energy materials for next generation Na-ion based energy storage technologies and their use in renewable energy systems" (Grant No.: DST/TMD/MES/2k16/77).

# Supplementary Information

**Role of porosity and diffusion coefficient in porous electrode used in supercapacitors- Correlating theoretical and experimental studies**

*Puja De[1], Joyanti Halder[1], Chinmayee Chowde Gowda[2], Sakshi Kansal[3], Surbhi Priya[3], Satvik Anshu[3], Ananya Chowdhury[1], Debabrata Mandal[2], Sudipta Biswas[1], Brajesh Kumar Dubey[4] and Amreesh Chandra[1, 2, 3]*

[1]Department of Physics, [2]School of Nanoscience and Technology, [3]School of Energy Science and Engineering, [4]Department of Civil Engineering, Indian Institute of Technology Kharagpur, Kharagpur, India-7213202.

*E-mail: achandra@phy.iitkgp.ac.in

**Synthesis of solid nanostructure of $Co_3O_4$ and $Mn_3O_4$:**

Solid nanostructure of $Co_3O_4$ and $Mn_3O_4$ were synthesized by following a typical precipitation method. In this method to synthesized $Co_3O_4$, first 0.05 M $Co(NO_3)_2$, $6H_2O$ was dissolved in 80 mL DI water at 40 °C and another 80 mL solvent of 0.1 M NaOH was prepared separately. Temperature of the solution was subsequently raised to 70 °C before dropwise adding into it the NaOH solution (80 ml). This led to the appearance of a light brown coloured precipitate. The overall colloidal solution was kept under vigorous stirring for 4 h. Following cooling back to room temperature, the precipitate was collected by centrifugation and washed three times each with DI water and methanol. The collected precipitate was vacuum dried at 60 °C for 12 h. The powder was finally calcined at 600 °C for 2 h in air.

To synthesized $Mn_3O_4$ solid structures, 0.22 molar of Manganese Chloride Tetrahydrate ($MnCl_2.4H_2O$), 1.37 mM hexadecyl trimethyl-ammonium Bromide (CTAB) was added under constant stirring. 0.44 M Sodium hydroxide (NaOH) solution was prepared separately and added dropwise to the first solution and was kept under constant stirring for 24 hours. The final precipitate was filtered and washed with DI water and ethanol 3 and 2 times respectively. The material was dried in the oven overnight at 80°C.

**Synthesis of solid structure of $Fe_2O_3$:**

For the formation of solid structure of $Fe_2O_3$, a solution of 0.4 M $FeSO_4 \cdot 7H_2O$ in water was added to an equal volume of 0.4 M oxalic acid solution. The solution was kept for stirring at room temperature and continued for 30 minutes. During the reaction, an iron (II) oxalate complex is formed which can be confirmed from the change of the colour of the solution from colourless to yellow. After completion, the product was centrifuged with DI followed by ethanol and subsequently left to dry overnight at 45 °C. The product was heated in presence of air at 400 °C with a slow ramp. During this process, the $OH^-$ group is removed and the iron (II) oxalate complex is converted to $Fe_2O_3$.

**Scheme of growth mechanism of porous structure materials:**

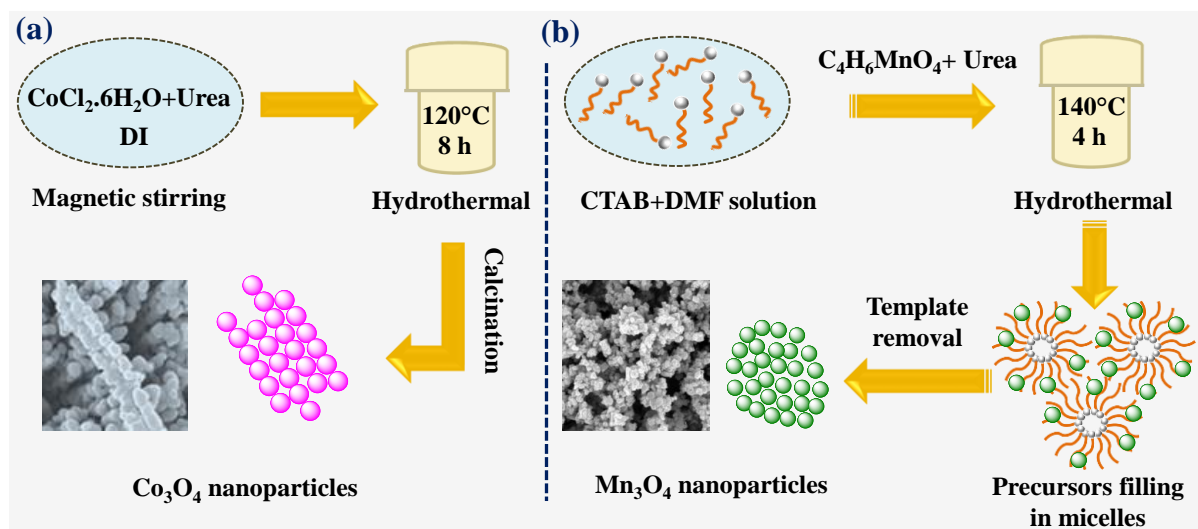

*Scheme S1: Growth mechanism of porous structure of (a) $Co_3O_4$ and (b) $Mn_3O_4$.*

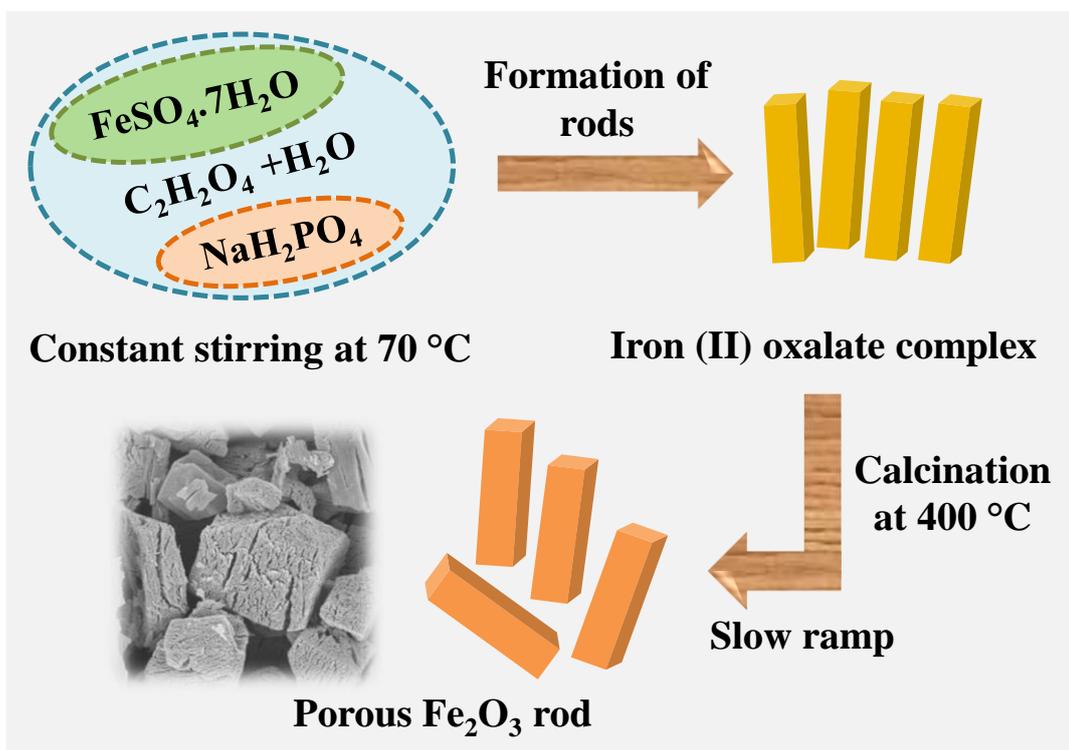

*Scheme S2: Growth mechanism of porous Fe₂O₃ rod.*

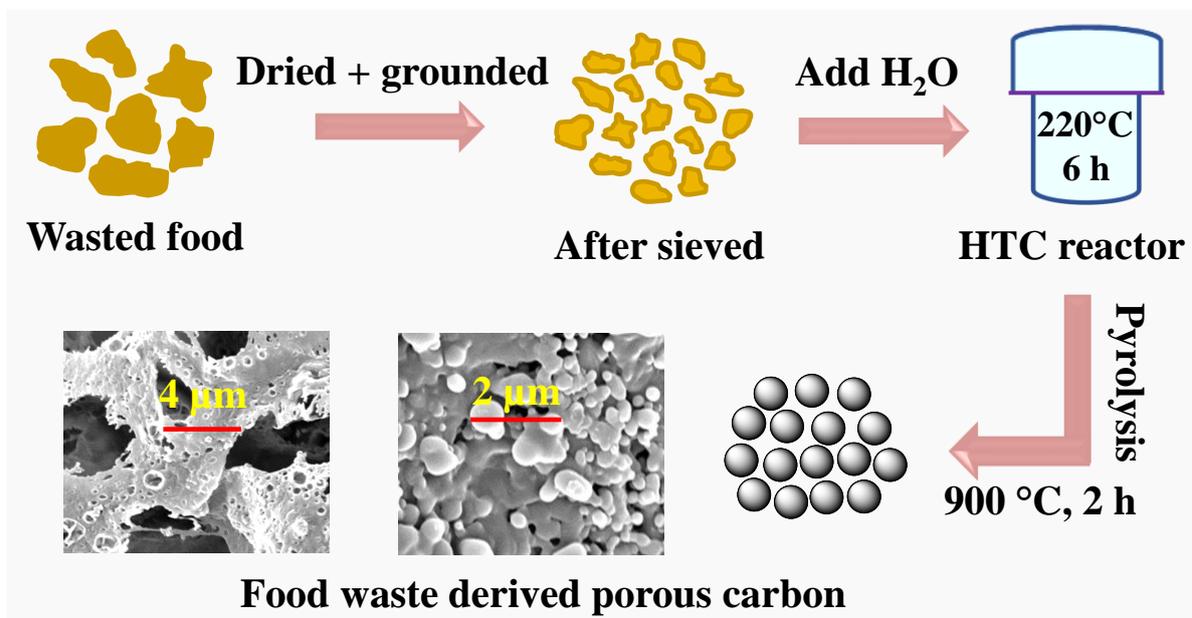

*Scheme S3: Growth mechanism of food waste derived porous carbon.*

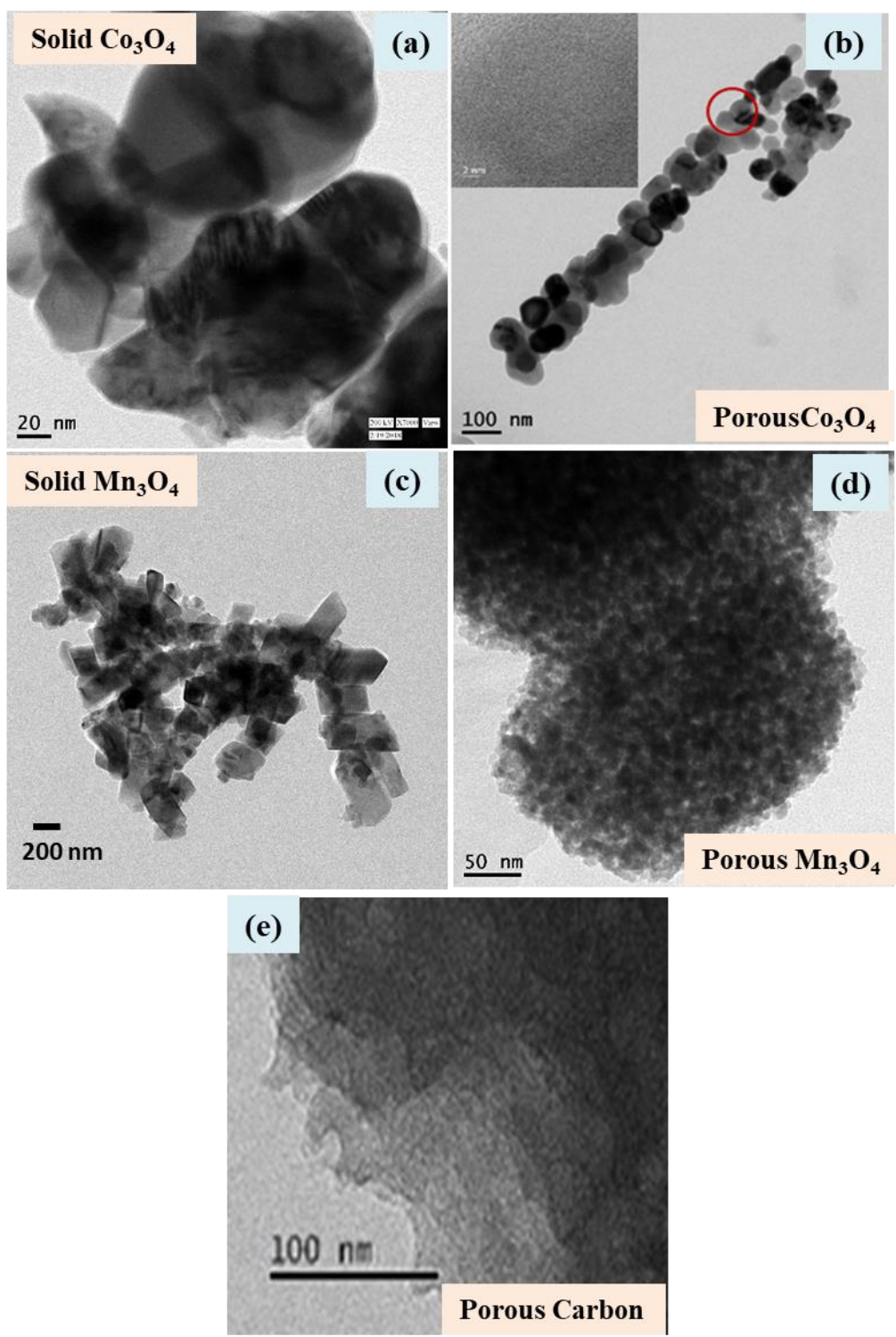

*Fig. S1: TEM micrographs of various materials*

*Table S1 Conductivity of various acidic, alkaline and neutral electrolytes*

| Types of aqueous electrolyte | Electrolyte | Conductivity (mS cm$^{-1}$) |
|---|---|---|
| Acidic electrolyte | Aqueous H$_2$SO$_4$ | 750 |
| Alkaline electrolyte | Aqueous KOH | 540 |
| | Aqueous NaOH | 221 |
| Neutral electrolyte | Aqueous KCl | 210 |
| | Aqueous NaCl | 90 |
| | Aqueous K$_2$SO$_4$ | 88.6 |
| | Aqueous Na$_2$SO$_4$ | 91.1 |

*Table S2 Hydration radius and ionic mobility of various electrolyte ions*

| Electrolyte Ion | Size of bare ion (Å) | Radius of hydration sphere (Å) | Ionic mobility (m$^2$ s$^{-1}$ V$^{-1}$) |
|---|---|---|---|
| H$^+$ | 1.15 | 2.80 | 36.2×10$^{-8}$ |
| K$^+$ | 1.33 | 3.31 | 7.6×10$^{-8}$ |
| Na$^+$ | 0.95 | 3.58 | 5.2×10$^{-8}$ |
| Li$^+$ | 0.60 | 3.82 | 3.9×10$^{-8}$ |
| OH$^-$ | 1.76 | 3.00 | 20.6×10$^{-8}$ |
| SO$_4^{2-}$ | 2.90 | 3.79 | 8.3×10$^{-8}$ |
| Cl$^-$ | 1.81 | 3.30 | 7.91×10$^{-8}$ |

*Equation S1: Formula to calculate specific capacitance from the cyclic voltammetry profile.*

$$C = \frac{1}{2mVs} \int_{-V}^{V} I \, dV$$

$C$ = specific capacitance,     $m$ = mass of the active material,     $V$ = potential window

$s$ = scan rate,     $\int_{-V}^{V} I \, dV$ = area under the CV curve

*Equation S2: Formula to calculate specific capacitance from the galvanostatic charge-discharge profile.*

$$C = \frac{I}{m} \frac{dt}{V - IR}$$

$C$ = specific capacitance,     $\frac{I}{m}$ = current density,     $dt$ = discharge time

$V$ = potential window,     $IR$ = voltage drop

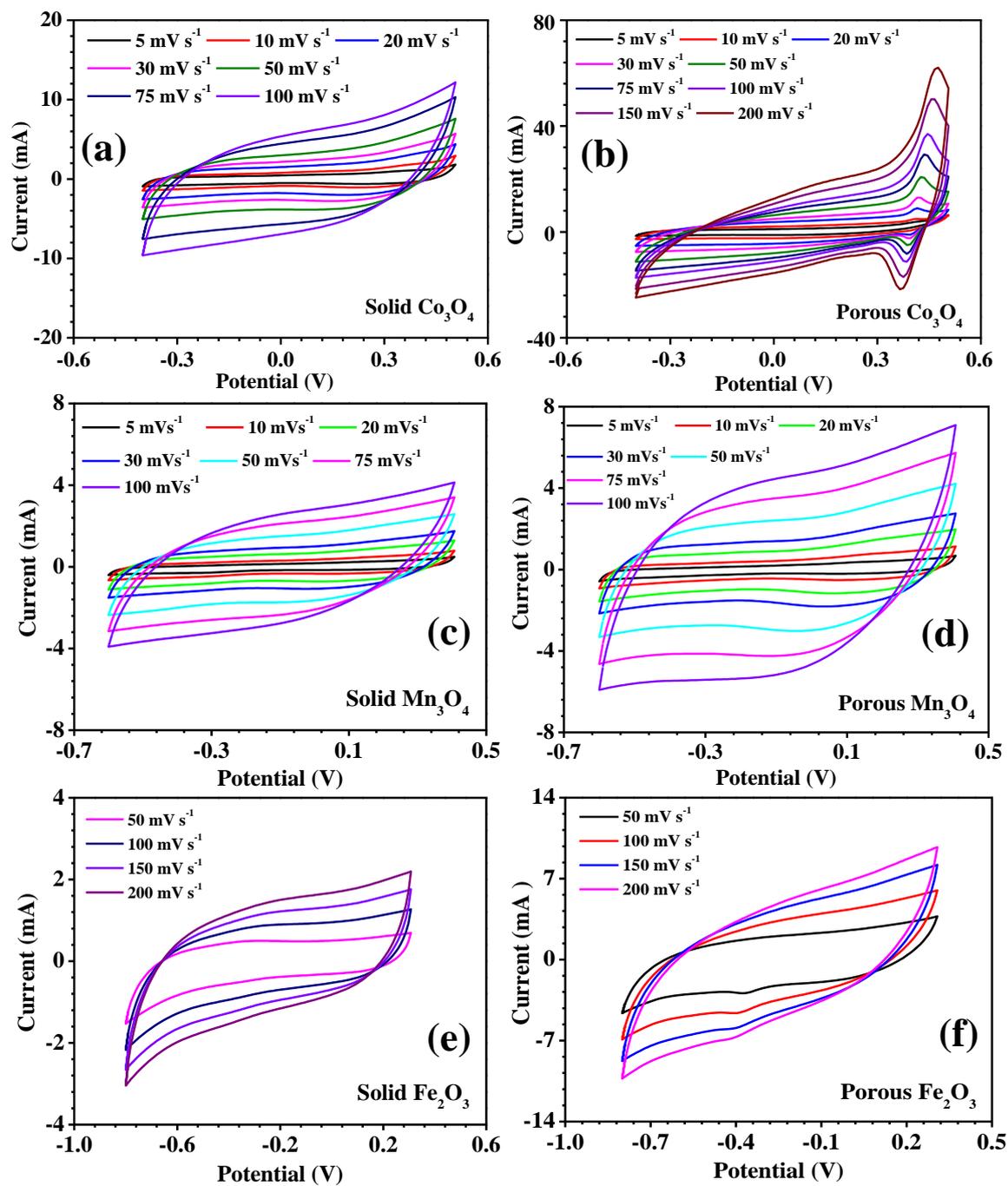

*Fig. S2: Cyclic voltammetry profile at various scan rates for solid and porous morphology of (a-b) $Co_3O_4$ (c-d) $Mn_3O_4$, and (e-f) $Fe_2O_3$.*

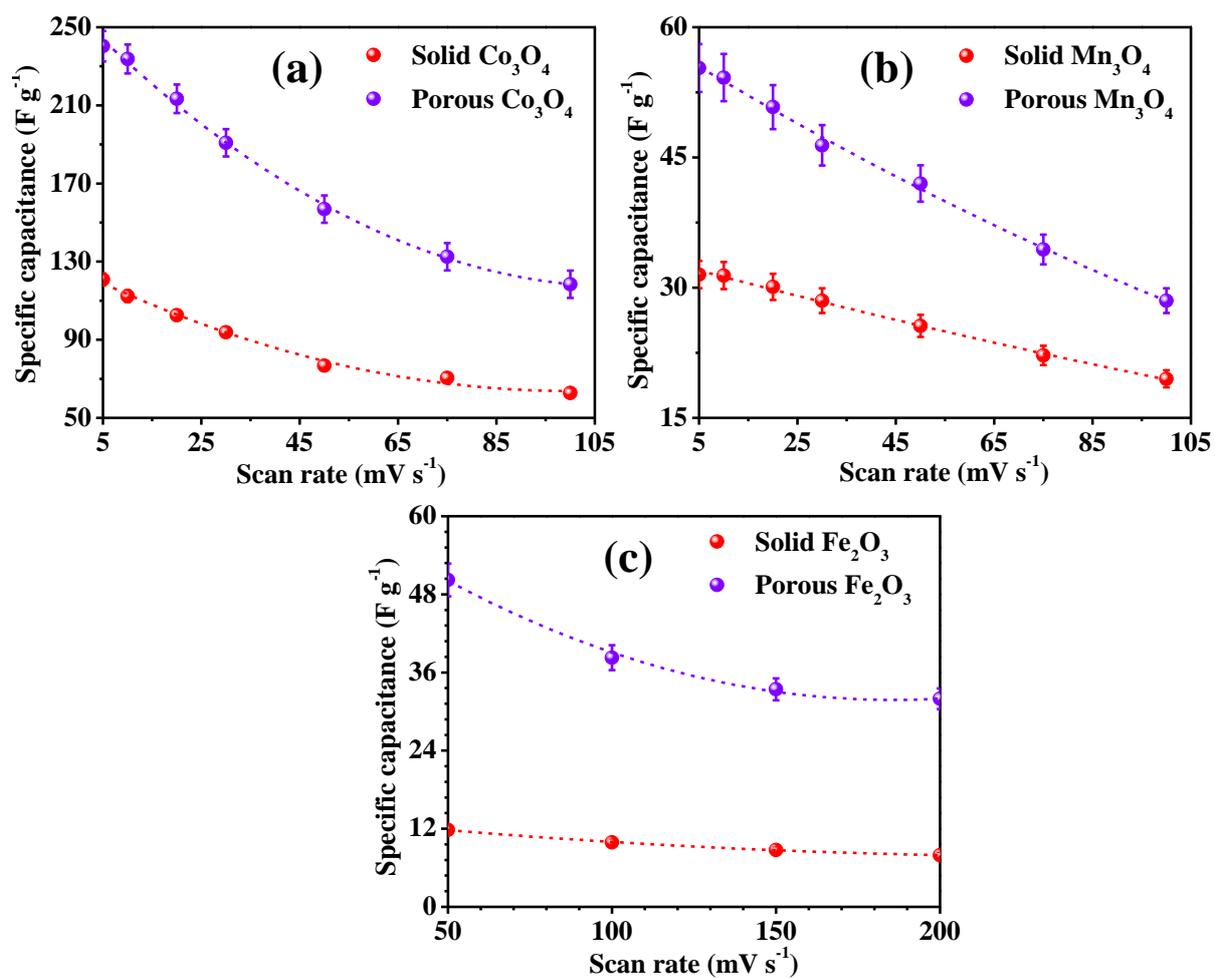

*Fig. S3: Variation of specific capacitance with scan rate for solid and porous morphology of (a) $Co_3O_4$ (b) $Mn_3O_4$, and (d) $Fe_2O_3$.*

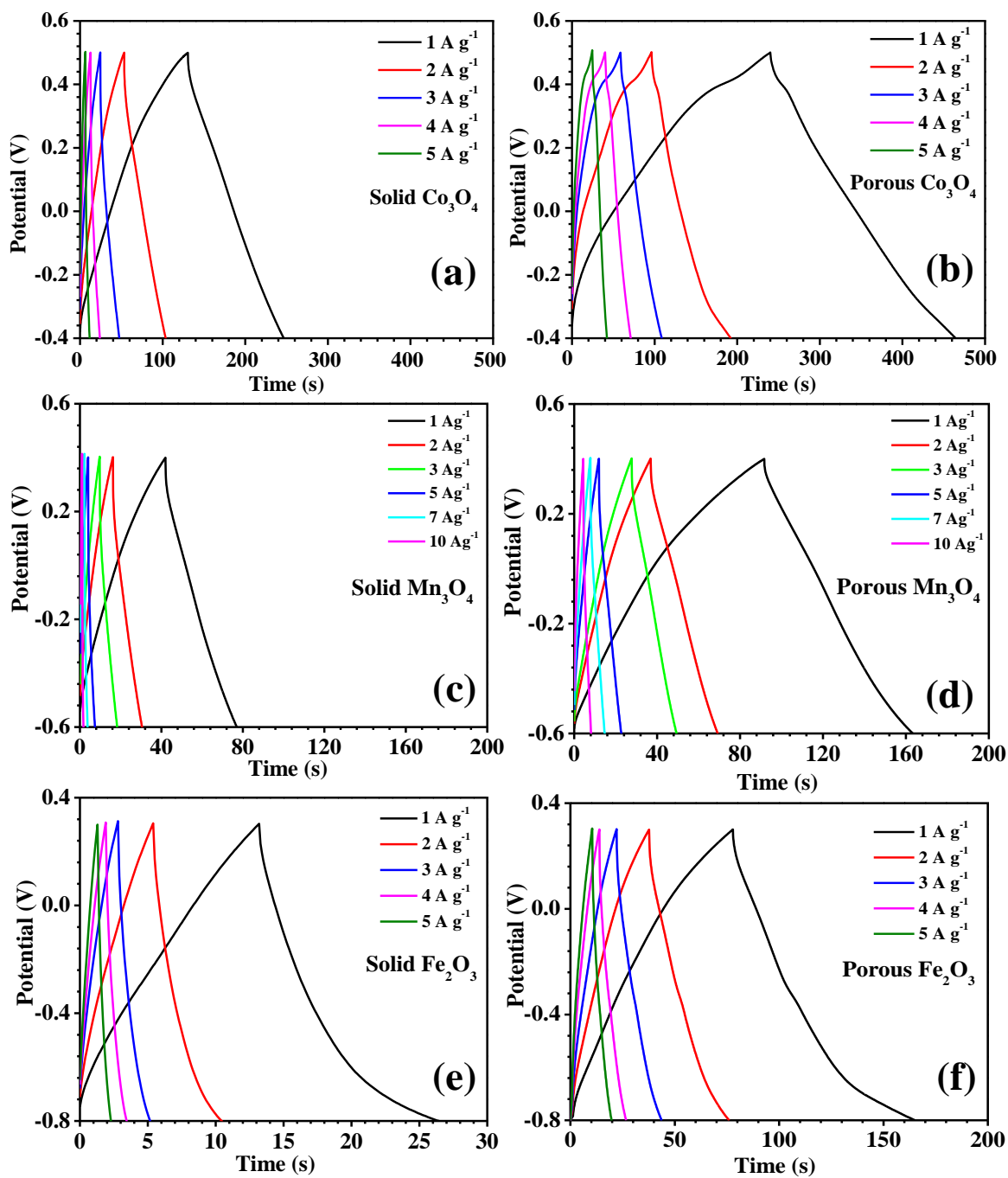

*Fig. S4: Charge-discharge profile at various current densities for solid and porous morphology of (a-b) $Co_3O_4$ (c-d) $Mn_3O_4$, and (e-f) $Fe_2O_3$.*

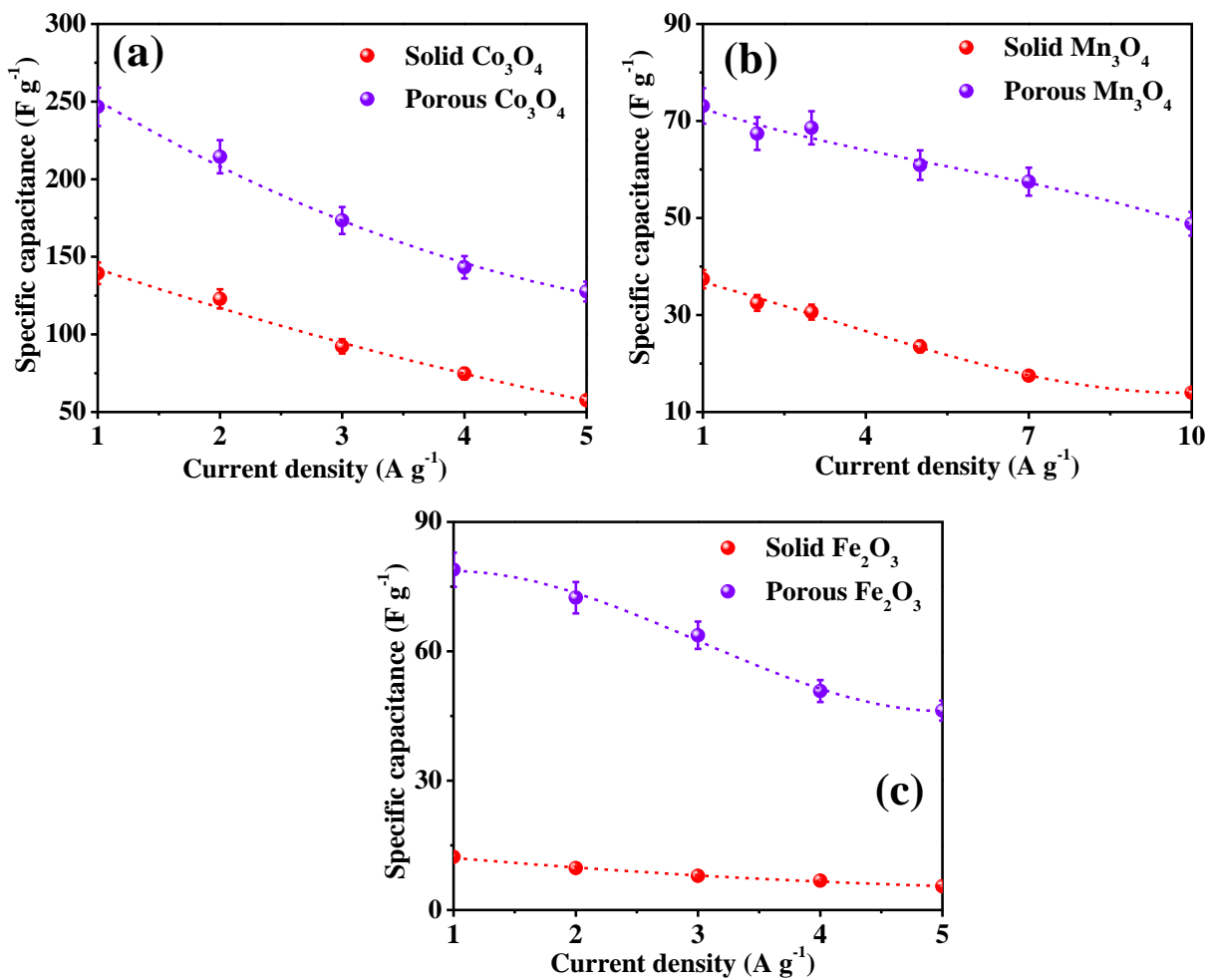

*Fig. S5: Variation of specific capacitance with current densities for solid and hollow morphology of (a) $Co_3O_4$ (b) $Mn_3O_4$, and (d) $Fe_2O_3$.*

**Electrochemical performances of solid and porous carbon:**

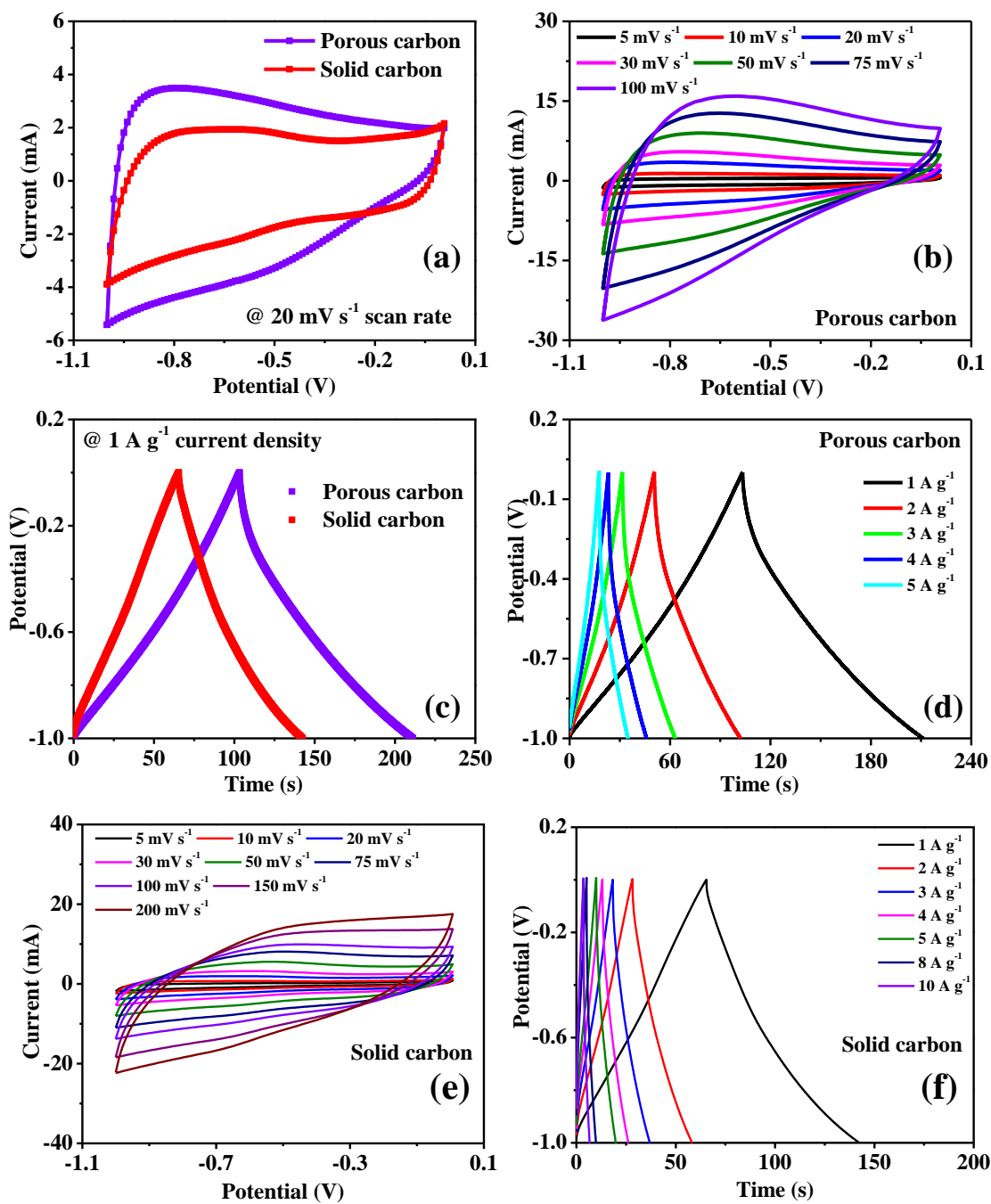

*Fig. S6: (a) Comparison of CV profile between solid and porous carbon at 20 mV s$^{-1}$ scan rate. (b) CV profile at various scan rates of porous carbon. (c) Compassion of CD profile at 1 A g$^{-1}$ current density between solid and porous structure of carbon. (d) CD profile at various current densities of porous carbon. (e) CV profile at various scan rates, and (f) CD profile at various current densities of solid carbon.*

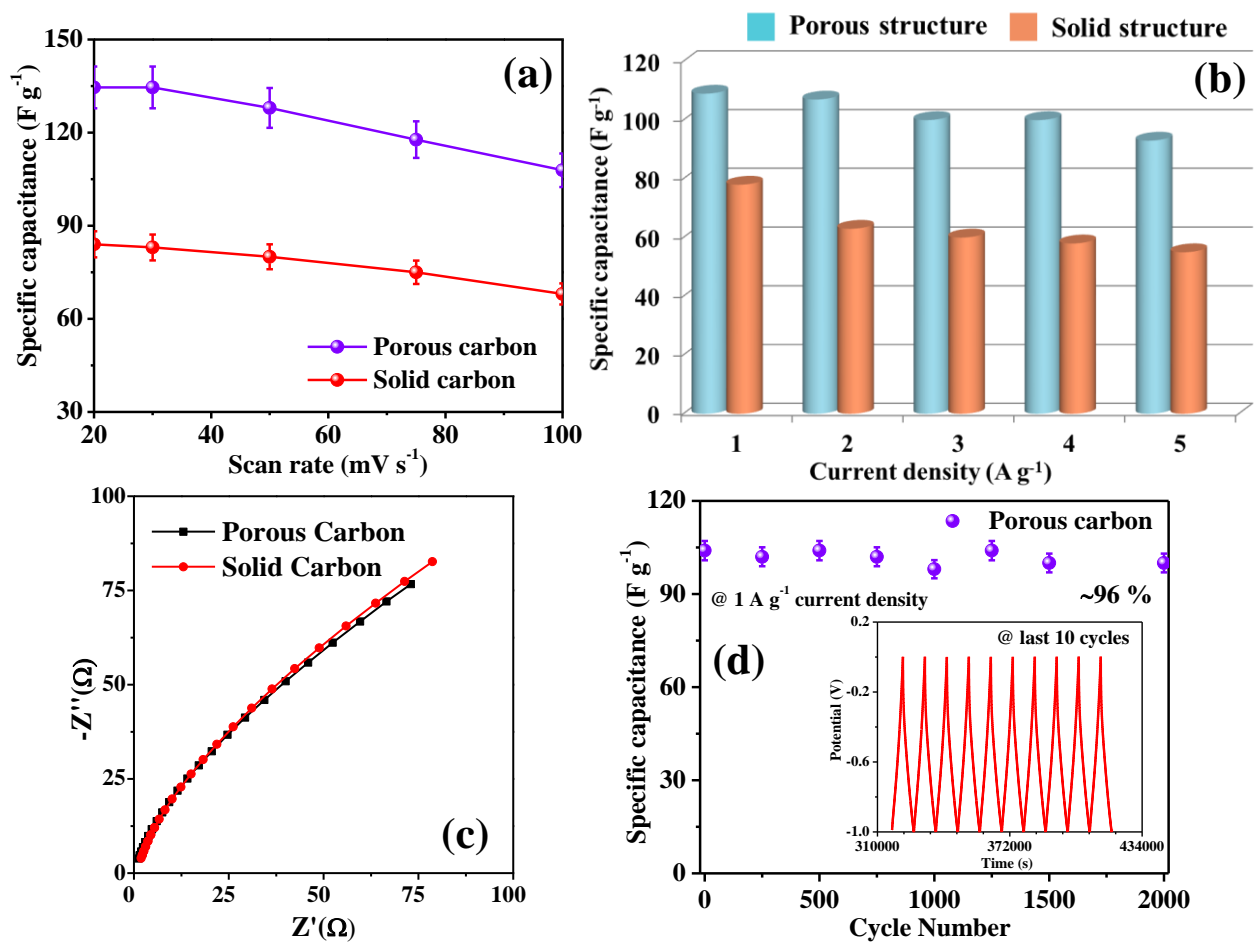

*Fig. S7: Variation of specific capacitance of solid and porous structure of carbon with (a) scan rates and (b) current densities. (c) Nyquist plot of solid and porous carbon and (c) cycling stability of food waste derived porous carbon.*

*Table S3: Cycling stability and charge-discharge efficiency of the porous structured materials.*

| Material | Structure | Cycling stability | Charge-discharge efficiency |
|---|---|---|---|
| $Co_3O_4$ | Porous | 97 % after 1,000 cycles | 83 % |
| $Mn_3O_4$ | Porous | 97 % after 1,000 cycles | 92 % |
| $Fe_2O_3$ | Porous | 98 % after 1,000 cycles | 82 % |
| Carbon | Porous | 96 % after 2,000 cycles | 94 % |

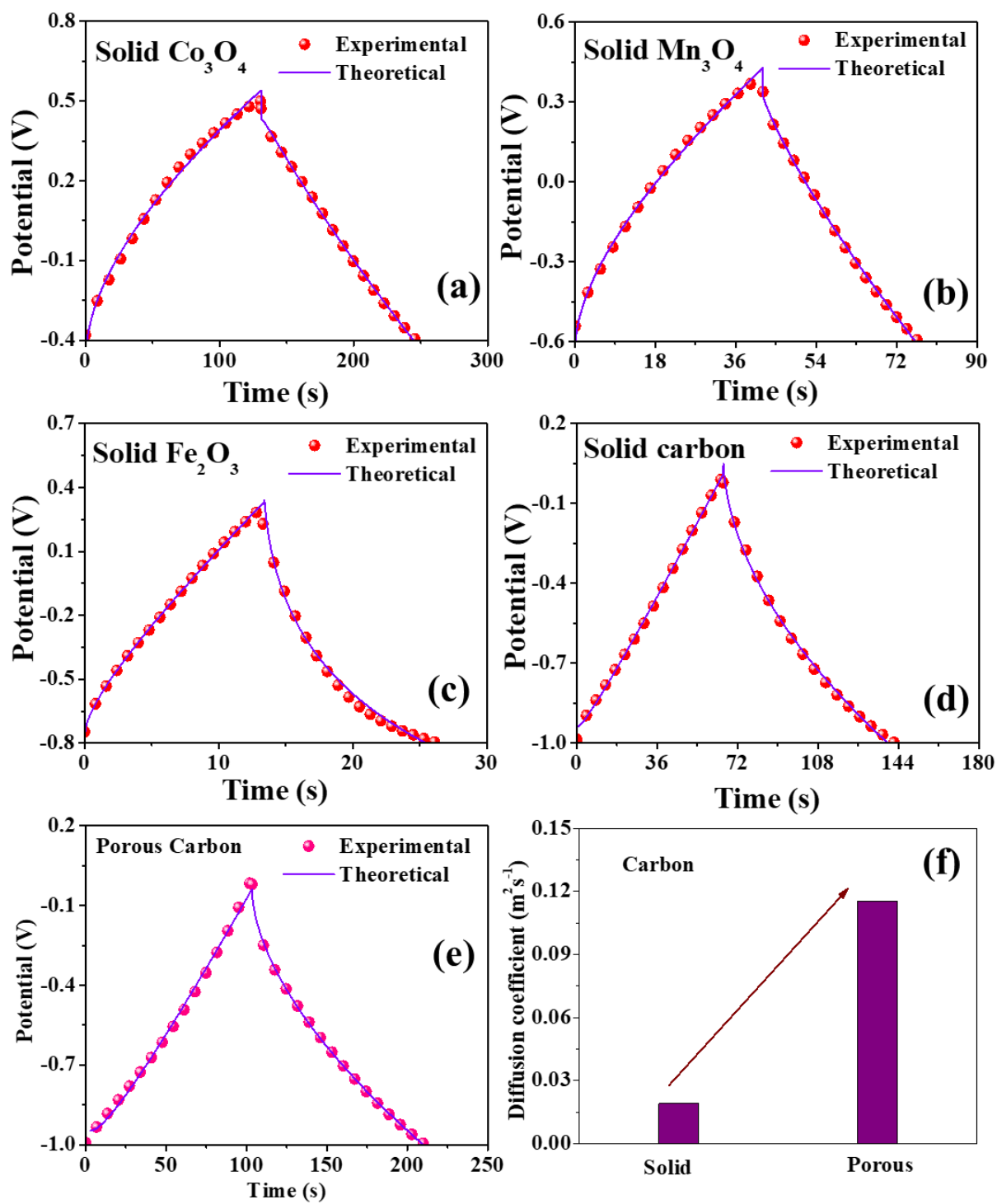

*Fig. S8: Comparison between theoretical and experimental data at 1 A g$^{-1}$ current density for solid structure of (a) Co$_3$O$_4$ (b) Mn$_3$O$_4$ (c) Fe$_2$O$_3$ (d) carbon, and (e) food waste derived porous carbon and (f) comparison of diffusion coefficient between solid and porous carbon.*

43